\documentclass[notitlepage,superscriptaddress,showpacs,nobalancelastpage,twocolumn,aps,prl,floatfix]{revtex4-1}
\usepackage[english]{babel}
\RequirePackage[T1]{fontenc}
\RequirePackage{times} 

\usepackage{siunitx} 
\usepackage[italicdiff]{physics}
\usepackage{amsfonts}
\usepackage{subfigure}
\usepackage{amsmath}
\usepackage{amssymb}
\usepackage{graphicx,epstopdf}
\usepackage{xspace}
\usepackage{array}
\usepackage[hidelinks]{hyperref}
\usepackage{color}
\usepackage{xcolor}
\usepackage{my_symbols}
\usepackage{CJK}
\usepackage{bm}
\usepackage{verbatim}
\usepackage{tikz}
\usepackage{float}
\usepackage{appendix}
\usepackage{pdfpages}

\makeatletter
\AtBeginDocument{\let\LS@rot\@undefined}
\makeatother
\usetikzlibrary{calc,3d}
\usetikzlibrary{arrows}
\usetikzlibrary{snakes}

\usepackage[normalem]{ulem}

{\par}
\newcounter{mycomment}

\newcommand\rmv{\bgroup\markoverwith {\textcolor{red}{\rule[0.5ex]{2pt}{0.4pt}}}\ULon}

\begin{document}
\begin{CJK*}{UTF8}{gbsn} 
\title{Dynamic Exchange Coupling between Magnets Mediated by Attenuating Elastic Waves}
\author{Weichao Yu (余伟超)}
\email{wcyu@fudan.edu.cn}
\affiliation{State Key Laboratory of Surface Physics and Institute for Nanoelectronic Devices and Quantum Computing, Fudan University, Shanghai 200433, China}
\affiliation{Zhangjiang Fudan International Innovation Center, Fudan University, Shanghai 201210, China}

\begin{abstract}
Coupling between spatially separated magnets can be mediated by excitations such as photons and phonons, which can be characterized as coherent coupling and dissipative coupling with real and imaginary coupling rate. We theoretically predict the existence of dynamic exchange coupling in a closed magneto-elastic system mediated by attenuating elastic waves and whose coupling rate is complex in general, leading to alternating repulsive or attractive spectrum depending on thickness of the elastic media. The presence of dynamic exchange coupling and its competition with coherent coupling are numerically verified according to the generalized Hooke's law in magneto-elastic systems. The predicted mechanism provides a new strategy to synchronize precessing magnets as well as other excitations over long distance and pave the way for non-Hermitian engineering of collective modes in hybrid magnonics, phononics and photonics.
\end{abstract}

\maketitle
\end{CJK*}

{\it Introduction.} Coupling between distant magnets is one of the research frontiers in the field of spintronics and magnonics \cite{barman_2021_2021}, by which the information carried by precessing magnetization can be efficiently transferred over long distance \cite{ruckriegel_long-range_2020}. At short range, magnets can be coupled to each other via direct coupling such as exchange interaction, Ruderman-Kittel-Kasuya-Yosida (RKKY) interaction \cite{liu_observation_2019} and indirect coupling via spin-pumping \cite{heinrich_dynamic_2003}. At longer range, it's natural to couple magnets directly via dipolar interaction and indirect coupling can be mediated by structured waves such as electromagnetic waves \cite{zare_rameshti_cavity_2022} and elastic waves \cite{bozhko_magnon-phonon_2020} in the presence of magnon-photon and magnon-phonon coupling.

\begin{figure}[b]
  \centering
  \includegraphics[width=8.5cm]{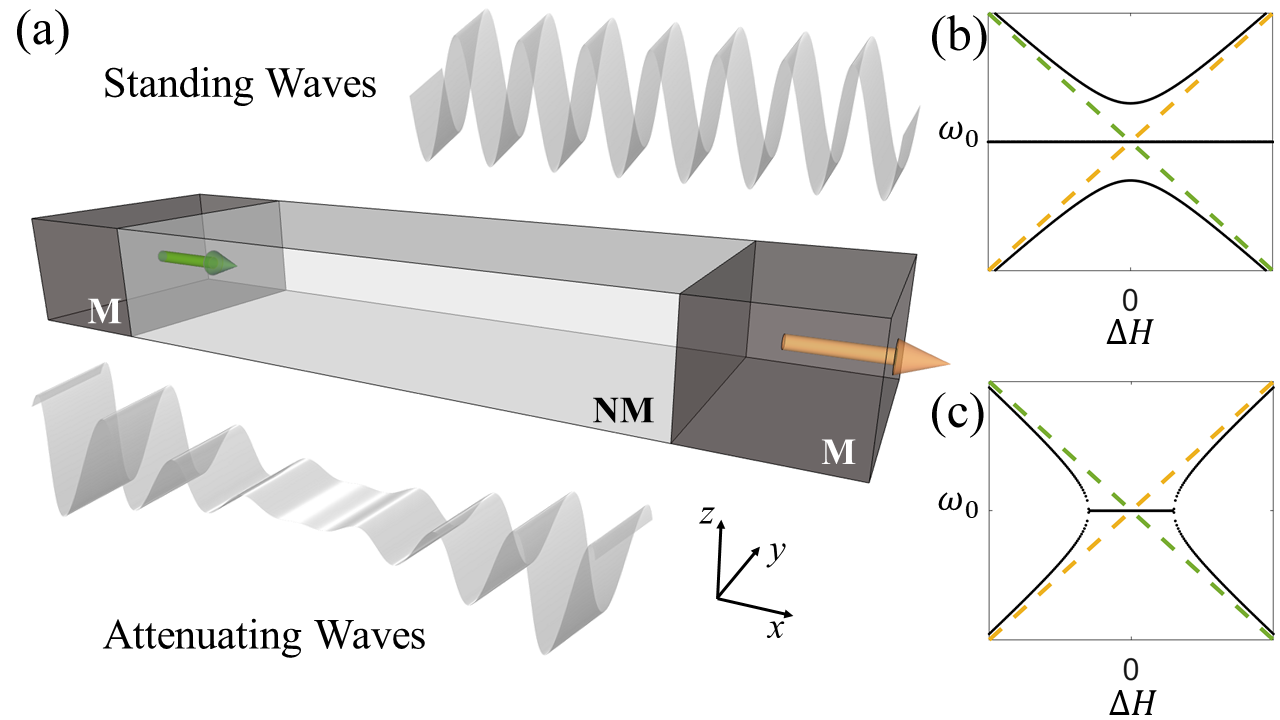}
  \caption{(a) Configuration of a closed tri-layer structure composed of a non-magnetic (NM) insulator sandwiched by two magnets (M). Two magnets couple with each other via elastic waves under field detuning $\Delta H$. (b) In the absence of acoustic attenuation, coupling between two magnets is mediated by standing elastic wave modes and repulsive spectrum is demonstrated at resonance as a consequence of coherent coupling. (c) Under same conditions as in (b) but in the presence of acoustic attenuation, coupling is mediated by attenuating elastic waves, resulting in attractive spectrum, produced by dynamic exchange coupling. Bare Kittel modes (dashed lines in orange and green) are plotted in (b) and (c) for eye guidance.}
  \label{fig1}
\end{figure}

There are mainly two types of features indicating coupling between magnets \cite{wang_dissipative_2020,harder_coherent_2021}: (i) avoided level crossings induced by coherent coupling, where two bare Kittel modes of magnets repel to each other (Fig.1(b)) and (ii) attractive levels induced by dissipative coupling or two-tone driving, where the levels attract to each other and the dynamics are synchronized (Fig.1(c)). In the past few years, coherent coupling between magnets have been widely studied in both magnon-photon \cite{zhang_magnon_2015,lambert_cavity-mediated_2016,zare_rameshti_indirect_2018,yu_circulating_2020,bourhill_generation_2023} and magnon-phonon \cite{berk_strongly_2019,an_coherent_2020,li_advances_2021,an_bright_2022} systems, while dissipative coupling (or two-tone driving) hasn't been reported in cavity magnonic systems until 2018 \cite{harder_level_2018,xu_cavity-mediated_2019,grigoryan_synchronized_2018,grigoryan_cavity-mediated_2019}.

In this Letter, we theoretically predict the existence of dynamic exchange coupling in a magneto-elastic system mediated by attenuating elastic waves, similar to the one realized by non-local pumping of spin currents \cite{heinrich_dynamic_2003} but from different physical origin. Different from coherent (dissipative) coupling with real (imaginary) coupling rate \cite{wang_dissipative_2020}, a distinct consequence of the dynamic exchange coupling predicted here is alternating spectrum with level repulsion and level attraction depending on the thickness of the elastic media. Consequently, collective modes of distant magnets can be manipulated in either synchronized or Rabi-like way \cite{zare_rameshti_cavity_2022}, since phonon decay length ($\sim$mm) is usually much larger than spin-diffusion length in normal metals ($\sim$nm) \cite{cornelissen_nonlocal_2017,ruckriegel_long-range_2020}.

{\it Phenomenological model.} Consider a closed tri-layer system of infinite lateral extent, composed of two magnets with thickness $d$ and saturated magnetization $M_\text{s}$ sandwiched by a non-magnetic insulator with thickness $L$, as shown in \Figure{fig1}(a). In the presence of magneto-elastic coupling, the precessing magnets pump phonons into the nonmagnetic insulator and at the same time, absorb phonons pumped from the other one. Hence the two magnets communicate with each other via phonon pumping \cite{streib_damping_2018} and the dynamics of unit magnetization vector $\mathbf{m}=\mathbf{M}/M_\text{s}$ is described by a set of coupled Landau-Lifshitz-Gilbert (LLG) equations \cite{heinrich_dynamic_2003}
\begin{equation}
\frac{\partial \mathbf{m}_\iota}{\partial t}=-\gamma\mathbf{m}_\iota \times \mathbf{H}_{\text{eff}}^\iota+(\alpha+\alpha^\prime)\mathbf{m}_\iota\times\frac{\partial \mathbf{m}_\iota}{\partial t}-\alpha^{\prime\prime}\mathbf{m}_{\bar{\iota}}\times\frac{\partial \mathbf{m}_{\bar{\iota}}}{\partial t},
\label{LLGdynamicexchange}
\end{equation}
where $\bar{\iota}=2, 1$ for $\iota=1, 2$, with gyromagnetic ratio $\gamma$, intrinsic Gilbert damping coefficient $\alpha$, effective damping coefficient $\alpha^\prime$  induced by phonon pumping and effective enhancement coefficient $\alpha^{\prime\prime}$ due to absorption of phonons. At this model-level stage, we focus on the regime of strong dissipation where the phonons reaching the opposite magnet experience attenuation, resulting into $\alpha^\prime>\alpha^{\prime\prime}$ and the dissipation is significant enough so that standing wave modes cannot be established. We consider effective field as $\mathbf{H}_{\text{eff}}^\iota=[H_0+(-1)^\iota\Delta H/2]\hat{\mathbf{x}}$ for perpendicular configuration, contributed by static external field $H_0$ and detuning field $\Delta H$ with dipolar effects and crystalline anisotropy disregarded for simplicity.

Decomposing the magnetization into a static part and a dynamical part, \textit{i.e.}, $\mathbf{m}_\iota=\mathbf{m}_\iota^0+\delta\mathbf{m}_\iota e^{i\omega t}$, and considering configuration of perpendicular magnetization $\mathbf{m}_\iota^0=\mathbf{m}_{\bar{\iota}}^0=(1,0,0)$, Eq.(\ref{LLGdynamicexchange}) can be linearized into frequency domain and reduced to
\begin{equation}
-\omega m_+^\iota+\left[\omega_0^\iota+i\omega\left(\alpha+\alpha^\prime\right)\right]m_+^\iota-i\alpha^{\prime\prime}\omega m_+^{\bar{\iota}}=0,
\label{CoupledEquation}
\end{equation}
with the definition of right-handed precessing mode $m_+^\iota=\delta m_y^\iota+i\delta m_z^\iota$ and $\omega_0^\iota=\gamma (H_0+(-1)^\iota\Delta H/2)$. Equation (\ref{CoupledEquation}) can be treated as the equation of motion of two coupled oscillators with state vector $|\psi\rangle=(m_+^\iota,m_+^{\bar{\iota}})$ governed by the Hamiltonian
\begin{equation}
\mathcal{H}=\left(
\begin{matrix}
-\omega+\omega_0^\iota+i\omega(\alpha+\alpha^\prime) & -i\alpha^{\prime\prime}\omega\\
-i\alpha^{\prime\prime}\omega &  -\omega+\omega_0^{\bar{\iota}}+i\omega(\alpha+\alpha^\prime)
 \end{matrix}\right).
\label{hamiltonian}
\end{equation}
The diagonal components of Eq.(\ref{hamiltonian}) determine bare frequencies of uncoupled Kittel modes while the non-diagonal components, which are complex in general, govern the dynamic exchange coupling between two magnets. It should be noted that real-valued coherent coupling is absent here since standing elastic waves are eliminated due to consideration of strong dissipation.

Phonon absorption is a reverse process of phonon pumping, and the two process will finally be balanced when the system reaches equilibrium ($\alpha^\prime=\alpha^{\prime\prime}$) in the absence of dissipation. However, in the current case, elastic waves are attenuated during propagation, leading to decrease of transverse momentum current \cite{streib_damping_2018} as well as a phase delay. We consider the effective enhancement coefficient in the form of $\alpha^{\prime\prime}=\alpha^\prime\text{exp}[-L/\Lambda+i2\pi L/\lambda]$, with decay length of phonons $\Lambda$ and wavelength of elastic waves $\lambda=2\pi c/\omega$ where $c$ is the transverse wave velocity. Hence we can define the non-diagonal terms in Eq.(\ref{hamiltonian}) (for perpendicular configuration) as a complex-valued coupling rate of dynamic exchange coupling
\begin{equation}
J_{(\perp)}=-i\omega\alpha^{\prime\prime}=-i\omega\alpha^\prime \text{exp}[-L/\Lambda+i2\pi L/\lambda].
\label{exchangeperp}
\end{equation}

\begin{figure}[h]
  \centering
  \includegraphics[width=8.8cm]{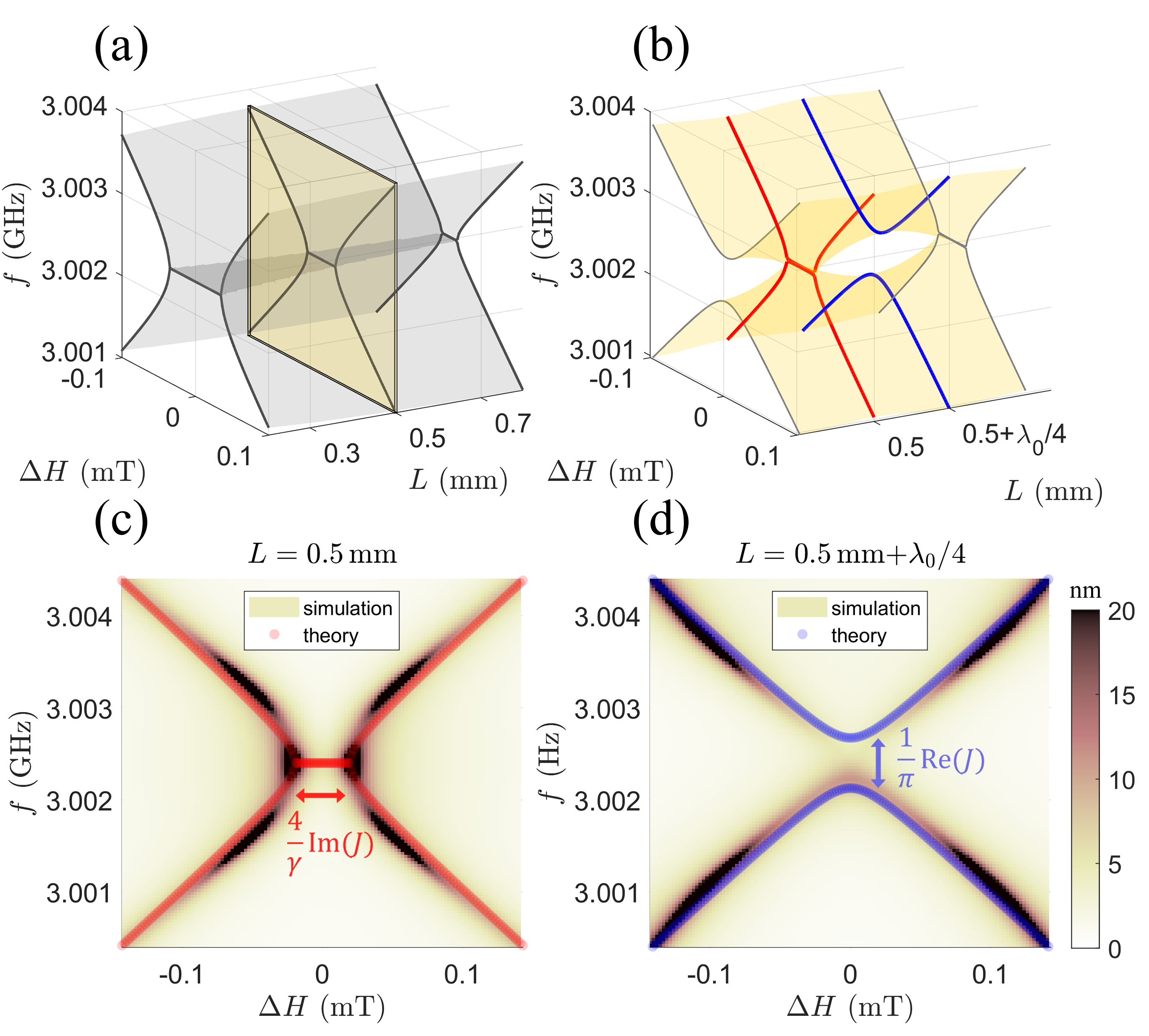}
  \caption{(a) Distance $L$ dependence of theoretical spectrum solved from Eq.(\ref{hamiltonian}) with damping induced by phonon pumping $\alpha^\prime=\,$\SI{2.5e-4}{} and amplified elastic damping $\beta=6\beta_0$. (b) Zoomed view of yellow box in (a), indicating that the spectrum shows either repulsive or attractive features depending on $L$ with period of half wavelength of elastic waves. Red and blue curves corresponds to the case of $L=\,$\SI{0.5}mm and $L=\,$\SI{0.5}{mm}$+\lambda_0/4$ with $\lambda_0=c_\text{NM}/f_0$ and $f_0=\,$\SI{3.0024}{GHz}, which are compared with spectrum of average phonon excitation $\langle|\mathbf{u}|\rangle$ extracted from numerical simulation in (c) and (d).}
  \label{fig2}
\end{figure}

Neglecting terms leading to local damping ($\alpha$ and $\alpha^\prime$) in diagonal components, the eigenstates of Eq.(\ref{hamiltonian}) when $\omega=\omega_0^\iota=\omega_0^{\bar{\iota}}$ are simplified to be
\begin{equation}
\Delta H=\pm 2 \text{Im}(J_{(\perp)})/\gamma,
\end{equation}
indicating two exceptional points whose distance corresponds to an effective exchange interaction competing with detuning field. The eigenfrequencies of Eq.(\ref{hamiltonian}) when $J$ is purely imaginary ($2L=n\lambda_0$ with $n$ a positive integer and $\lambda_0=2\pi c/\omega_0$) are plotted in \Figure{fig1}(c), demonstrating a typical spectrum of level attraction. On the other hand, when $J$ is purely real ($2L=(n+\frac{1}{2})\lambda_0$), the levels repel to each other with splitting spectral distance $2J_{(\perp)}/(2\pi)$, similar to coherent coupling. The dependence of spectrum on distance $L$ between two magnets is shown in \Figure{fig2}(a). The magnitude of coupling strength $J$ exponentially decays with $L$ due to attenuation of elastic waves, and at the same time, the spectrum alternates between level repulsion and level attraction with period of $\lambda_0/2$, as seen in \Figure{fig2}(b). The effect of impedance mismatch and thickness of magnetic layer $d$ are disregarded in this minimal model.

{\it Generalized Hooke's law for numerical simulation.} In order to verify the validity of the model, we establish a numerical approach, which is essentially generalized Hooke's law in the presence of magneto-elastic coupling. The energy density of magneto-elastic coupling for cubic system is given by \cite{kittel_interaction_1958}
\begin{equation}
\begin{split}
\begin{aligned}
\mathcal{U}_\text{me}&=b_1\sum\limits_{i}m_i^2\varepsilon_{ii}+b_2\sum\limits_{i\neq j}m_im_j\varepsilon_{ij},
\label{mecoupling}
\end{aligned}
\end{split}
\end{equation}
where $b_1$ and $b_2$ are magneto-elastic constants and strain tensor is defined as $\varepsilon_{ij}=\left(\partial u_i/\partial x_j+\partial u_j/\partial x_i\right)/2$ with lattice displacement vector $\mathbf{u}=(u_x,u_y,u_z)$. Here the unit magnetization vector is expressed in global coordinate as $\mathbf{m}=(m_x,m_y,m_z)$ whose static component $\mathbf{m}_0=(\sin\theta_0\cos\phi_0,\sin\theta_0\sin\phi_0,\cos\theta_0)$ with $\theta_0$ and $\phi_0$ the polar and azimuthal angle. For arbitrary static configuration, expression in global coordinate can be transformed to local coordinate according to $(m_x,m_y,m_z)^T=\mathcal{R}(\theta_0,\phi_0)(m_1,m_2,m_3)^T$ with $\mathcal{R}(\theta_0,\phi_0)$ the rotation matrix \footnote[1]{$\mathcal{R}(\theta_0,\phi_0)=\left(
 \begin{smallmatrix}
 \cos\theta_0\cos\phi_0 & -\sin\phi_0 & \sin\theta_0\cos\phi_0 \\
  \cos\theta_0\sin\phi_0 & \cos\phi_0 & \sin\theta_0\sin\phi_0  \\
  -\sin\theta_0 & 0 & \cos\theta_0
 \end{smallmatrix}\right)$.}, so that the decomposition in local coordinate becomes $\mathbf{m}\simeq\mathbf{m}_0+\delta\mathbf{m}e^{i\omega t}=(0,0,1)+(m_1,m_2,0)e^{i\omega t}$ and $m_1, m_2 \ll 1$ according to linear approximation \cite{dreher_surface_2012}. Effective dynamic field induced by magneto-elastic coupling can be derived as
 \begin{equation}
\delta\mathbf{h}=-\frac{1}{\mu_0M_\text{s}}\frac{\partial \mathcal{U}_\text{me}}{\partial(\delta\mathbf{m})},\label{effectivefield}
\end{equation}
which is also expressed in local coordinate $\delta\mathbf{h}=(h_1, h_2, 0)$ with vacuum permeability $\mu_0$. Along with applied external field $\mathbf{H}=\mathbf{H}_0+\delta\mathbf{H}e^{i\omega t}$ with $\mathbf{H}_0=(0,0,H_0)$ and $\delta\mathbf{H}=(H_1,H_2,0)$, the linear response of magnetization perturbed by effective field is given by
\begin{equation}
M_\text{s}\delta m_i=\chi_{ij}(\delta H_j+\delta h_j),
\label{linearresponse}
\end{equation}
where $\chi_{ij}$ is the \emph{Polder susceptibility tensor} \cite{stancil_spin_2009} whose components are
\begin{subequations}
\begin{align}
&\chi_{11}=\chi_{22}=\frac{\omega_\text{M}(\omega_0-i\omega\alpha)}{(\omega_0-i\omega\alpha)^2-\omega^2},\\
&\chi_{12}=-\chi_{21}=\frac{-i\omega\omega_\text{M}}{(\omega_0-i\omega\alpha)^2-\omega^2},
\end{align}
\label{susceptibilitytensor}
\end{subequations}
with $\omega_0=\gamma (H_0\pm \Delta H/2)$ and $\omega_\text{M}=\gamma M_\text{s}$. Equation (\ref{linearresponse}) can be further rewritten as
\begin{equation}
M_\text{s}\delta m_i=\widetilde\chi_{ij}(\delta \widetilde{H}_j+\delta \widetilde{h}_j),
\label{linearresponse2}
\end{equation}
by defining $\widetilde\chi_{ij}=\chi_{ij}/(\mu_0M_\text{s}^2)$, $\delta\widetilde{\mathbf{H}}=\mu_0M_\text{s}\delta\mathbf{H}$ and $\delta\widetilde{\mathbf{h}}=\mu_0M_\text{s}\delta\mathbf{h}$, so that $\delta\widetilde{\mathbf{H}}$ and $\delta\widetilde{\mathbf{h}}$ are in dimension of \emph{effective stress}. Plugging Eq.(\ref{linearresponse2}) back to Eq.(\ref{mecoupling}), we obtain an explicit form of energy density $\mathcal{U}_\text{me}(\theta_0,\phi_0,\varepsilon_{ij},\delta\mathbf{H})$ including backaction of magnetic excitation by strain as well as applied field.

According to theory of linear elasticity \cite{rinaldi_theory_1985,luthi_physical_2007,dreher_surface_2012}, the effective stress can be obtained according to $\sigma_{ij}^\text{me}=\partial \mathcal{U}_\text{me}/\partial\varepsilon_{ij}$. Combining Eq.(\ref{mecoupling},\ref{effectivefield},\ref{linearresponse2}), neglecting high-order terms such as $\varepsilon^2$ and $\chi^2$, and following the convention of \emph{Voigt notation} that $\sigma_i=(\sigma_{xx},\sigma_{yy},\sigma_{zz},\sigma_{yz},\sigma_{xz},\sigma_{xy})$ and $\varepsilon_i=(\varepsilon_{xx},\varepsilon_{yy},\varepsilon_{zz},2\varepsilon_{yz},2\varepsilon_{xz},2\varepsilon_{xy})$, one can obtain the generalized Hooke's law in the presence of magneto-elastic coupling (see Supplemental Materials (SM) \cite{SM}),
\begin{equation}
\sigma_{i}^\text{me}=\frac{\partial \mathcal{U}_\text{me}}{\partial\varepsilon_{i}}=C_{ij}^\text{me}\varepsilon_{j}+\sigma_{i}^\text{H},
\label{hookeslaw}
\end{equation}
where $C_{ij}^\text{me}=\partial^2\mathcal{U}_\text{me}/(\partial\varepsilon_i\partial\varepsilon_j)$ is a second-rank stiffness tensor contributed by magneto-elastic coupling, and $\sigma_{i}^\text{H}$ is effective stress induced by external driving field. There is an extra term $\sigma_i^0$ not shown in Eq.(\ref{hookeslaw}), which leads to magnetostriction induced by static magnetization $\mathbf{m}_0$  \cite{sato_dynamic_2021} and is eliminated since it doesn't contribute to dynamics in frequency domain. All terms in Eq.(\ref{hookeslaw}) are derived explicitly in Supplemental Materials \cite{SM}.

With the generalized Hooke's law, we are able to simulate the hybrid magneto-elastic system by solving the equation of motion for elastic waves in frequency domain
\begin{equation}
\rho\omega^2\mathbf{u}-i\beta\omega\mathbf{u}+\nabla\cdot\bar{\boldsymbol{\sigma}}_\text{tot}=0,
\label{equationofmotionfrequencydomain}
\end{equation}
with vector of displacement field $\mathbf{u}(\mathbf{r},t)=\mathbf{u}(\mathbf{r})e^{i\omega t}$, where $\rho$ and $\beta$ are mass density and elastic damping coefficient for specific materials. Components of the total stress tensor $\bar{\boldsymbol{\sigma}}_\text{tot}$ for magnetic materials are $\sigma_i^\text{tot}=\sigma_i^\text{me}+C_{ij}^\text{el}\varepsilon_j$, with $C_{ij}^\text{el}$ the elasticity tensor. For a cubic system, $C_{11}^\text{el}=C_{22}^\text{el}=C_{33}^\text{el}=2\mu+\lambda$, $C_{12,21}^\text{el}=C_{13,31}^\text{el}=C_{23,32}^\text{el}=\lambda$ and $C_{44}^\text{el}=C_{55}^\text{el}=C_{66}^\text{el}=\mu$, with $\lambda$ and $\mu$ the L\'{a}me constants.

The equation of motion Eq.(\ref{equationofmotionfrequencydomain}) is numerically solved by COMSOL Multiphysics \cite{comsol} based on finite-element method. We simulate the tri-layer structure in \Figure{fig1}(a) by considering two typical materials, \textit{i.e.}, yttrium iron garnet (YIG) for magnets (M) and gadolinium gallium garnet (GGG) for nonmagnets (NM), known for high acoustic quality and widely used in experiments \cite{an_coherent_2020,an_bright_2022,schlitz_magnetization_2022}. For YIG \cite{clark_elastic_1961}, $\rho_\text{M}=\,$\SI{5170}{kg/m^3}, $\mu_\text{M}=\,$\SI{7.64e10}{J/m^3}, $\lambda_\text{M}=\,$\SI{1.16e11}{J/m^3}, $\gamma=\,$\SI{2.21e5}{Hz/(A/m)}, $M_\text{s}=\,$\SI{1760}{Oe}, $\alpha=\,$\SI{1.3e-4}{} and thickness for both layers $d=\,$\SI{200}{nm}. For GGG \cite{kleszczewski_phononphonon_1988}, $\rho_\text{NM}=\,$\SI{7070}{kg/m^3}, $\mu_\text{NM}=\,$\SI{9e10}{J/m^3}, $\lambda_\text{NM}=\,$\SI{1.11e11}{J/m^3}. Magneto-elastic constants $b_2=2b_1=\,$\SI{6.96e5}{J/m^3} for YIG \cite{comstock_magnetoelastic_1965}. Phonon relaxation rate in GGG is measured as $\eta(f)/(2\pi)=(144\text{[kHz]}+5.2\times10^{-6}\text{[1/GHz]}f^2)$ \cite{schlitz_magnetization_2022}, and we consider identical elastic damping coefficient for both YIG and GGG as $\beta_0=2\eta\rho_\text{NM}$ for simplicity \cite{an_coherent_2020}, leading to decay length in GGG $\Lambda=\rho c_\text{NM}/\beta_0\simeq\,$\SI{3}{mm} at $f=\,$\SI{3}{GHz} with $c_\text{NM}=\sqrt{\mu_\text{NM}/\rho_\text{NM}}=\,$\SI{3568}{m/s}.

In the simulation, both of the two magnets are excited by a right-handed field in same phase, leading to an effective stress $\widetilde{H}_1=-i\widetilde{H}_2=\,$\SI{1e4}{Pa}. Spectrum is produced by sweeping average phonon excitation $\langle|\mathbf{u}|\rangle=(1/V)\int\sqrt{|u_x|^2+|u_y|^2+|u_z|^2}\,dV$ with $V$ the volume of the whole structure. It's shown in \Figure{fig2}(c) and (d) that the simulation results agree well with the theoretical prediction with the fitting parameter $\alpha^\prime=\,$\SI{2.5e-4}{}. A rigorous theoretical estimation of $\alpha^\prime$ will be discussed later.

\begin{figure}[t]
  \centering
  \includegraphics[width=8.5cm]{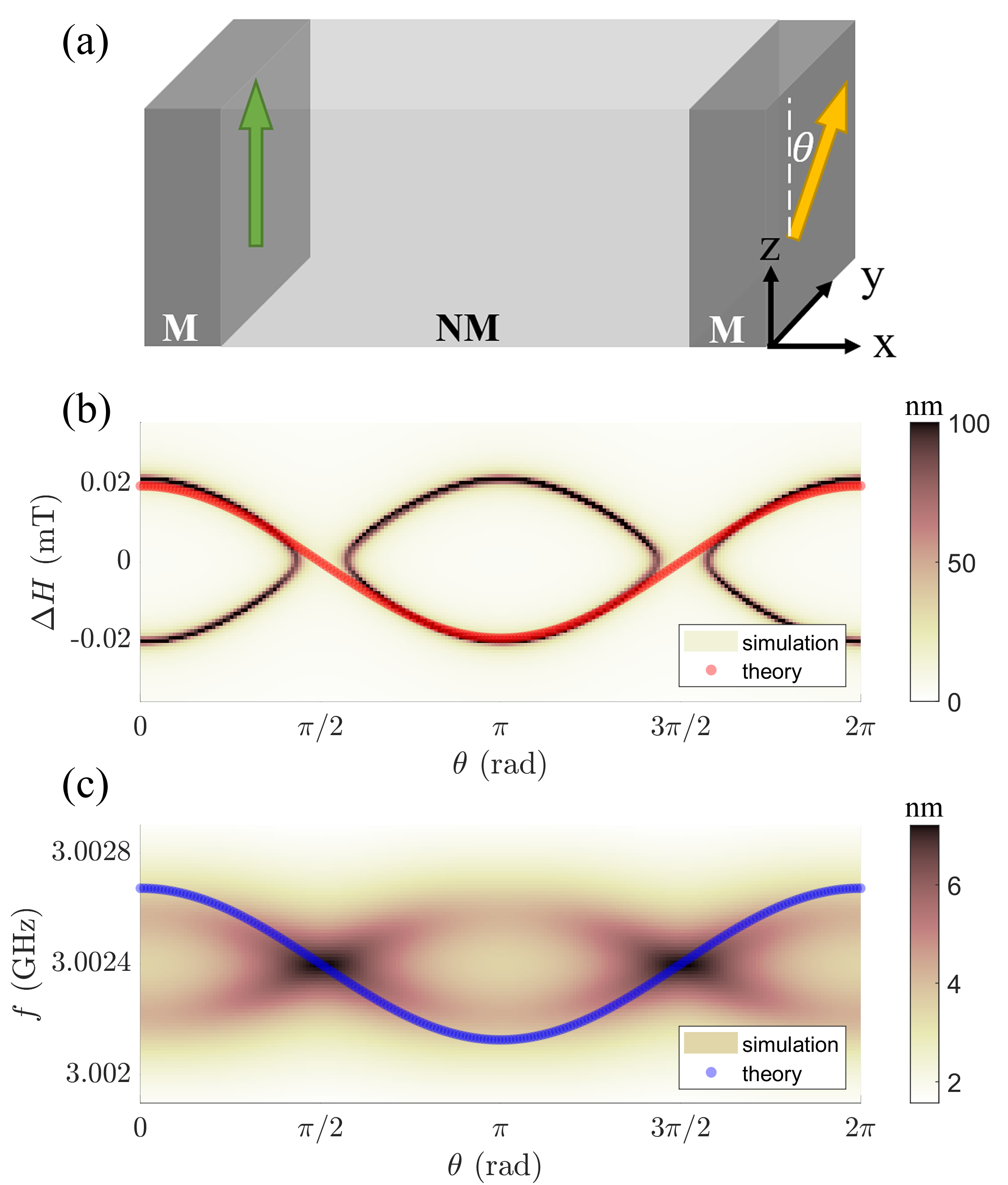}
  \caption{(a) Schematics for in-plane magnetized configuration with relative angle $\theta$. Angular dependence of simulated spectrum with central frequency $f_0=\,$\SI{3.0024}{GHz} for (b) $L=\,$\SI{0.5}{mm} with detuning and (c) $L=\,$\SI{0.5}{mm}$+\lambda_0/4$ without detuning. Red and blue curves are plotted according to theoretical prediction Eq.(\ref{exchangeperp}) and Eq.(\ref{exchangepara}) for $\alpha^\prime=\,$\SI{2.5e-4}{}. Amplified elastic damping $\beta=6\beta_0$ is considered.}
  \label{fig3}
\end{figure}

{\it Dynamic exchange coupling for in-plane configuration.} We further investigate the in-plane configuration where two magnets are both magnetized and detuned in y-z plane with relative angle $\theta$, as in \Figure{fig3}(a). Simulation is performed by applying a right-handed field on one of the magnets, keeping the other one passively excited via dynamic exchange coupling. Same as in \Figure{fig2}, two cases are studied for $L=\,$\SI{0.5}{mm} ($J$ is purely imaginary) and $L=\,$\SI{0.5}{mm}$+\lambda_0/4$ ($J$ is purely real). As shown in \Figure{fig3}(b) and (c), it's numerically confirmed that the dynamic exchange coupling rate for in-plane configuration shows a $\cos\theta$ dependence and can be expressed as
\begin{equation}
J_{(\parallel)}=J_{(\perp)}\mathbf{m}_0^\iota\cdot\mathbf{m}_0^{\bar{\iota}},
\label{exchangepara}
\end{equation}
which results into an effective detuning $\Delta H=\pm 2\text{Im}(J_{(\parallel)})/\gamma$ (red curve in \Figure{fig3}(b) for positive branch) and splitting levels $f=f_0\pm J_{(\parallel)}/(2\pi)$ (blue curve in \Figure{fig3}(c) for positive branch), respectively for two cases. This is expected since a magnet for in-plane configuration pumps transverse phonons whose polarization is along the equilibrium magnetization, namely $\mathbf{u}^\iota\parallel\mathbf{m}_0^\iota$ \cite{streib_damping_2018}, resulting in a coupling rate $\sim\mathbf{u}^\iota\cdot\mathbf{m}_0^{\bar{\iota}}$ corresponding to a factor of $\cos\theta$. It has been calculated by Streib \textit{et al.} \cite{streib_damping_2018} that phonon pumping is always less efficient for the in-plane configuration. Generally speaking, for arbitrary magnetization configuration, the phenomenological parameter $\alpha^\prime$ (as well as $\alpha^{\prime\prime}$) in Eq.(\ref{LLGdynamicexchange}) is diagonal element of a second-rank tensor $\bar{\boldsymbol{\alpha}}^\prime$ whose trace corresponds to total Gilbert damping enhancement by phonon pumping. For the specific case in this work without crystalline anisotropy, we have the relation $\text{tr}(\bar{\boldsymbol{\alpha}}^\prime_{(\parallel)})=(1/2)\text{tr}(\bar{\boldsymbol{\alpha}}^\prime_{(\perp)})$ \footnote[2]{For example, $\bar{\boldsymbol{\alpha}}^\prime_{(\perp)}=\left(
 \begin{smallmatrix}
 0 &  &  \\
   & \alpha^\prime &   \\
   &  & \alpha^\prime
 \end{smallmatrix}\right)$ for perpendicular magnetization along $\hat{\mathbf{x}}$ and $\bar{\boldsymbol{\alpha}}^\prime_{(\parallel)}=\left(
 \begin{smallmatrix}
 0 &  &  \\
   & \alpha^\prime &   \\
   &  & 0
 \end{smallmatrix}\right)$ for in-plane magnetization along $\hat{\mathbf{y}}$}, consistent with the theoretical prediction \cite{streib_damping_2018}, thus leading to identical magnitude for dynamic exchange coupling rate $|J_{(\parallel)}|=|J_{(\perp)}|$ for collinear configurations.

\begin{figure}[b]
  \centering
  \includegraphics[width=7.5cm]{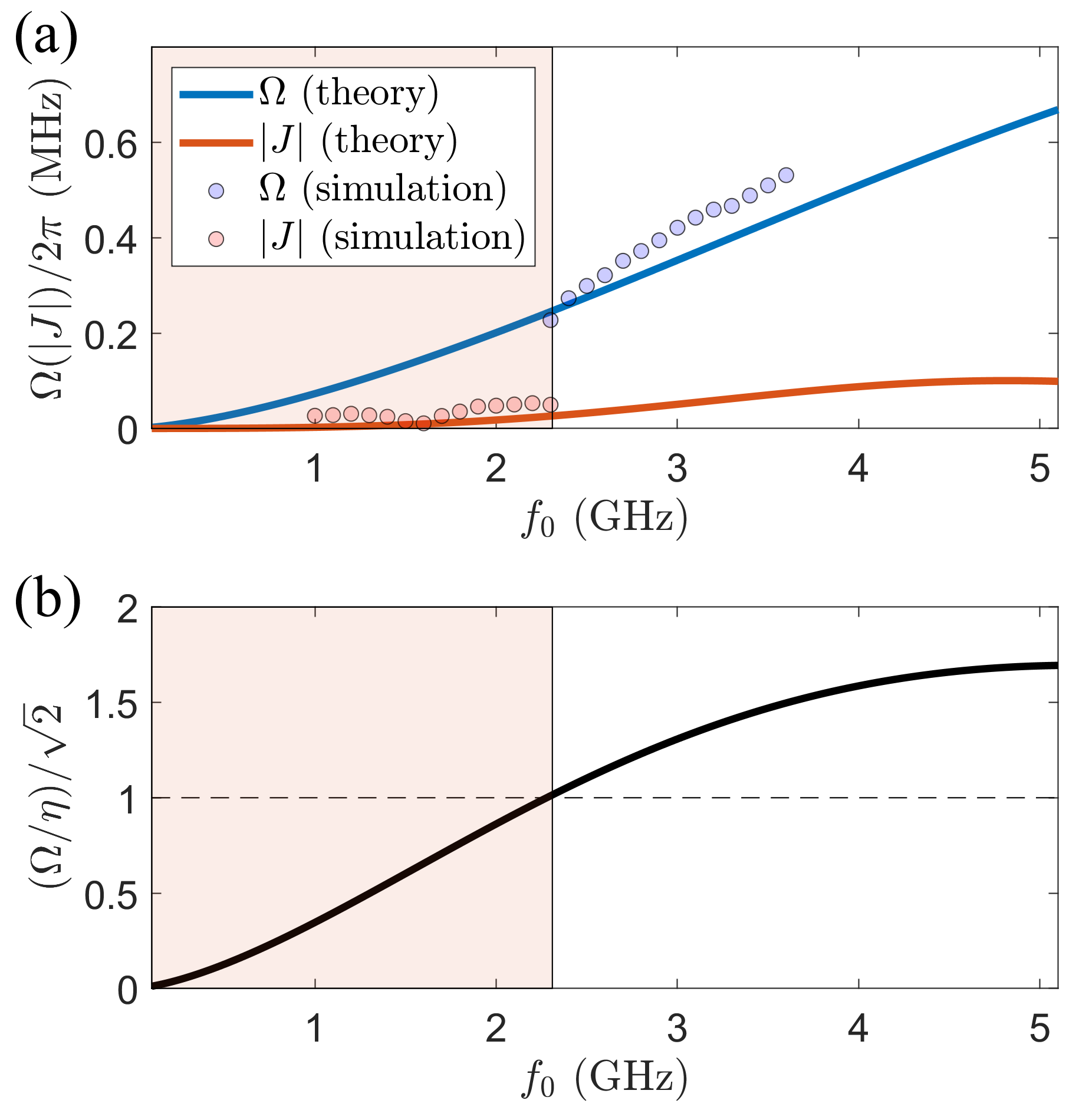}
  \caption{(a) Frequency dependence of coherent coupling strength $\Omega$ (blue) and magnitude of dynamic exchange coupling strength $|J|$ when $2L=n\lambda$ (red). Original elastic damping $\beta=\beta_0$ of GGG with thickness $L=\,$\SI{5}{mm} is considered. Circles are extracted from simulation and curves are plotted according to theoretical calculation. (b) Frequency dependence of relative strength between the coherent coupling rate and phonon relaxation rate $\Omega/(\sqrt{2}\eta)$, separating the weak coupling regime ($\Omega/(\sqrt{2}\eta)<1$) where dynamic exchange coupling dominates and strong coupling regime ($\Omega/(\sqrt{2}\eta)>1$) where coherent coupling dominates.}
  \label{fig4}
\end{figure}

{\it Discussion.} All calculations above are assuming amplified elastic damping coefficient $\beta=6\beta_0$, leading to phonon decay length $\Lambda\sim\,$\SI{0.5}{mm}, comparable to the thickness of GGG, so that dynamic exchange coupling dominates. In reality, GGG is an excellent conductor of phononic currents with $\Lambda\sim\,$\SI{2}{mm} (for $\beta=\beta_0$ and $f=\,$\SI{5}{GHz}) \cite{an_coherent_2020,an_bright_2022}, much larger than the thickness adopted in experiments, so that only coherent coupling has been reported in literatures so far. Based on the understanding above, we predict that it's possible to observe dynamic exchange coupling using existing experimental setups \cite{an_coherent_2020,an_bright_2022,an_optimizing_2023} by simply increasing thickness of GGG spacer. Figure 4 shows simulation results for perpendicular configuration with $L=\,$\SI{5}{mm} and realistic elastic damping $\beta=\beta_0$. In this case, coherent coupling is no longer negligible, whose strength is denoted as $\Omega$, which determines splitting magnitude of repulsive levels and has been calculated analytically \cite{an_coherent_2020,litvinenko_tunable_2021,schlitz_magnetization_2022,SM}. Theoretical strength of dynamic exchange coupling can be calculated according to Eq.(\ref{exchangeperp}) with $\alpha^\prime$ derived analytically in \cite{streib_damping_2018,SM}. Both coupling strength are extracted from simulation (red and blue circles) and compared with theoretical curves in \Figure{fig4}(a). We find that (i) Even for non-magnetic spacer with high acoustic quality, it's possible to observe dominant dynamic exchange coupling, with criteria that $\Omega/(\sqrt{2}\eta)<1$ \footnote[3]{$\Omega$ is defined as indirect coherent coupling between two magnets meadiated by standing elastic waves, hence $\Omega/\sqrt{2}$ indicates direct coupling strength between single magnet and the standing wave mode.}, corresponding to weak (coherent) coupling regime \cite{zhang_strongly_2014}, as indicated in \Figure{fig4}(b). Above the threshold (\SI{2.3}{GHz} in this case), the system moves to strong coupling regime and coherent coupling dominates. Around the transition region, both features of level attraction and level repulsion are present \cite{SM}. Therefore, it's indicated that dynamic exchange coupling and coherent coupling are a set of competing mechanism from same physical origin, \textit{i.e.}, elastic waves mediation, but approaching from different dissipation limit. (ii) Simulated strength for both coherent coupling and dynamic exchange coupling are larger than the theoretical ones. This is consistent with experimental measurements \cite{schlitz_magnetization_2022}, because inhomogeneous contributions are disregarded in theoretical model but preserved in simulation.

 It has been revealed that the dynamic exchange coupling mediated by attenuating elastic waves is closely related to dissipative coupling in cavity magnonic systems \cite{wang_nonreciprocity_2019,wang_dissipative_2020,zare_rameshti_cavity_2022}. Although dissipation is needed to wipe out coherent coupling, the dynamic exchange coupling is distinct from dissipative coupling, and is more conceptually close to two-tone driving \cite{grigoryan_synchronized_2018}, where the phase delay is realized by the propagation of elastic waves rather than an extra circuit element. On the other hand, the tri-layer structure in this model can be treated as a phononic cavity, and the results can be qualitatively reproduced by an effective three-oscillator model (two Kittel modes of magnets mediated by one standing elastic wave mode with strong dissipation) \cite{yu_prediction_2019}. However, the three-oscillator model cannot capture the microscopic mechanism of the complex-valued coupling rate. Although the phononic cavity is closed, the attenuation of elastic waves plays as an open channel which dissipates energy to an auxiliary reservoir, putting intrinsic dissipation \cite{hu_auxiliary_nodate} and open environment \cite{wang_unpublished_2023} on equal footing to realize level attraction.

{\it Conclusion.} We demonstrate an example of realizing dynamic exchange coupling mediated by attenuating elastic waves in a closed phononic cavity, which provides a new strategy to synchronize dynamics of magnets over long distance. A phenomenological model is proposed to reveal the wave nature of the coupling mechanism and is verified by numerical simulations. The proposed mechanism also offers a new way to engineer complex coupling strength between magnets. Along with chiral coupling between magnons and surface acoustic waves \cite{xu_nonreciprocal_2020,yamamoto_non-reciprocal_2020,yu_nonreciprocal_2020} and manipulation of local gain and loss \cite{christensen_parity-time_2016}, one is armed with a full non-Hermitian toolbox to study macroscopic spin chain with peculiar dispersions and Weyl criticality \cite{tserkovnyak_exceptional_2020}, which is promising for straintronic applications \cite{bandyopadhyay_magnetic_2021}. Although demonstrated in magneto-elastic system, the concept can be generalized to other coupled systems such as cavity magnonics \cite{zare_rameshti_cavity_2022}, optomechanics \cite{zhang_dissipative_2022} and hybrid quantum systems where propagating electromagnetic waves can be attenuated in materials with dielectric loss, so that multiple objects such as magnets, resonators and qubits can be either synchronized or desynchronized with attracted or repelled energy levels.

{\it Acknowledgements.}
I thank Shengtao Fan, Takuma Sato, Kei Yamamoto, Gerrit E. W. Bauer, Jiongjie Wang and Jiang Xiao for fruitful discussions. This work was supported by National Natural Science Foundation of China (Grant No. 12204107), Shanghai Pujiang Program (Grant No. 21PJ1401500) and Shanghai Science and Technology Committee (Grant No. 21JC1406200).


\begin{thebibliography}{52}%
\makeatletter
\providecommand \@ifxundefined [1]{%
 \@ifx{#1\undefined}
}%
\providecommand \@ifnum [1]{%
 \ifnum #1\expandafter \@firstoftwo
 \else \expandafter \@secondoftwo
 \fi
}%
\providecommand \@ifx [1]{%
 \ifx #1\expandafter \@firstoftwo
 \else \expandafter \@secondoftwo
 \fi
}%
\providecommand \natexlab [1]{#1}%
\providecommand \enquote  [1]{``#1''}%
\providecommand \bibnamefont  [1]{#1}%
\providecommand \bibfnamefont [1]{#1}%
\providecommand \citenamefont [1]{#1}%
\providecommand \href@noop [0]{\@secondoftwo}%
\providecommand \href [0]{\begingroup \@sanitize@url \@href}%
\providecommand \@href[1]{\@@startlink{#1}\@@href}%
\providecommand \@@href[1]{\endgroup#1\@@endlink}%
\providecommand \@sanitize@url [0]{\catcode `\\12\catcode `\$12\catcode
  `\&12\catcode `\#12\catcode `\^12\catcode `\_12\catcode `\%12\relax}%
\providecommand \@@startlink[1]{}%
\providecommand \@@endlink[0]{}%
\providecommand \url  [0]{\begingroup\@sanitize@url \@url }%
\providecommand \@url [1]{\endgroup\@href {#1}{\urlprefix }}%
\providecommand \urlprefix  [0]{URL }%
\providecommand \Eprint [0]{\href }%
\providecommand \doibase [0]{http://dx.doi.org/}%
\providecommand \selectlanguage [0]{\@gobble}%
\providecommand \bibinfo  [0]{\@secondoftwo}%
\providecommand \bibfield  [0]{\@secondoftwo}%
\providecommand \translation [1]{[#1]}%
\providecommand \BibitemOpen [0]{}%
\providecommand \bibitemStop [0]{}%
\providecommand \bibitemNoStop [0]{.\EOS\space}%
\providecommand \EOS [0]{\spacefactor3000\relax}%
\providecommand \BibitemShut  [1]{\csname bibitem#1\endcsname}%
\let\auto@bib@innerbib\@empty
\bibitem [{\citenamefont {Barman}\ \emph {et~al.}(2021)\citenamefont {Barman},
  \citenamefont {Gubbiotti}, \citenamefont {Ladak}, \citenamefont {Adeyeye},
  \citenamefont {Krawczyk}, \citenamefont {Gr\"{a}fe}, \citenamefont
  {Adelmann}, \citenamefont {Cotofana}, \citenamefont {Naeemi}, \citenamefont
  {Vasyuchka}, \citenamefont {Hillebrands}, \citenamefont {Nikitov},
  \citenamefont {Yu}, \citenamefont {Grundler}, \citenamefont {Sadovnikov},
  \citenamefont {Grachev}, \citenamefont {Sheshukova}, \citenamefont
  {Duquesne}, \citenamefont {Marangolo}, \citenamefont {Csaba}, \citenamefont
  {Porod}, \citenamefont {Demidov}, \citenamefont {Urazhdin}, \citenamefont
  {Demokritov}, \citenamefont {Albisetti}, \citenamefont {Petti}, \citenamefont
  {Bertacco}, \citenamefont {Schultheiss}, \citenamefont {Kruglyak},
  \citenamefont {Poimanov}, \citenamefont {Sahoo}, \citenamefont {Sinha},
  \citenamefont {Yang}, \citenamefont {M\"{u}nzenberg}, \citenamefont
  {Moriyama}, \citenamefont {Mizukami}, \citenamefont {Landeros}, \citenamefont
  {Gallardo}, \citenamefont {Carlotti}, \citenamefont {Kim}, \citenamefont
  {Stamps}, \citenamefont {Camley}, \citenamefont {Rana}, \citenamefont
  {Otani}, \citenamefont {Yu}, \citenamefont {Yu}, \citenamefont {Bauer},
  \citenamefont {Back}, \citenamefont {Uhrig}, \citenamefont {Dobrovolskiy},
  \citenamefont {Budinska}, \citenamefont {Qin}, \citenamefont {Dijken},
  \citenamefont {Chumak}, \citenamefont {Khitun}, \citenamefont {Nikonov},
  \citenamefont {Young}, \citenamefont {Zingsem},\ and\ \citenamefont
  {Winklhofer}}]{barman_2021_2021}%
  \BibitemOpen
  \bibfield  {author} {\bibinfo {author} {\bibfnamefont {A.}~\bibnamefont
  {Barman}}, \bibinfo {author} {\bibfnamefont {G.}~\bibnamefont {Gubbiotti}},
  \bibinfo {author} {\bibfnamefont {S.}~\bibnamefont {Ladak}}, \bibinfo
  {author} {\bibfnamefont {A.~O.}\ \bibnamefont {Adeyeye}}, \bibinfo {author}
  {\bibfnamefont {M.}~\bibnamefont {Krawczyk}}, \bibinfo {author}
  {\bibfnamefont {J.}~\bibnamefont {Gr\"{a}fe}}, \bibinfo {author}
  {\bibfnamefont {C.}~\bibnamefont {Adelmann}}, \bibinfo {author}
  {\bibfnamefont {S.}~\bibnamefont {Cotofana}}, \bibinfo {author}
  {\bibfnamefont {A.}~\bibnamefont {Naeemi}}, \bibinfo {author} {\bibfnamefont
  {V.~I.}\ \bibnamefont {Vasyuchka}}, \bibinfo {author} {\bibfnamefont
  {B.}~\bibnamefont {Hillebrands}}, \bibinfo {author} {\bibfnamefont {S.~A.}\
  \bibnamefont {Nikitov}}, \bibinfo {author} {\bibfnamefont {H.}~\bibnamefont
  {Yu}}, \bibinfo {author} {\bibfnamefont {D.}~\bibnamefont {Grundler}},
  \bibinfo {author} {\bibfnamefont {A.~V.}\ \bibnamefont {Sadovnikov}},
  \bibinfo {author} {\bibfnamefont {A.~A.}\ \bibnamefont {Grachev}}, \bibinfo
  {author} {\bibfnamefont {S.~E.}\ \bibnamefont {Sheshukova}}, \bibinfo
  {author} {\bibfnamefont {J.-Y.}\ \bibnamefont {Duquesne}}, \bibinfo {author}
  {\bibfnamefont {M.}~\bibnamefont {Marangolo}}, \bibinfo {author}
  {\bibfnamefont {G.}~\bibnamefont {Csaba}}, \bibinfo {author} {\bibfnamefont
  {W.}~\bibnamefont {Porod}}, \bibinfo {author} {\bibfnamefont {V.~E.}\
  \bibnamefont {Demidov}}, \bibinfo {author} {\bibfnamefont {S.}~\bibnamefont
  {Urazhdin}}, \bibinfo {author} {\bibfnamefont {S.~O.}\ \bibnamefont
  {Demokritov}}, \bibinfo {author} {\bibfnamefont {E.}~\bibnamefont
  {Albisetti}}, \bibinfo {author} {\bibfnamefont {D.}~\bibnamefont {Petti}},
  \bibinfo {author} {\bibfnamefont {R.}~\bibnamefont {Bertacco}}, \bibinfo
  {author} {\bibfnamefont {H.}~\bibnamefont {Schultheiss}}, \bibinfo {author}
  {\bibfnamefont {V.~V.}\ \bibnamefont {Kruglyak}}, \bibinfo {author}
  {\bibfnamefont {V.~D.}\ \bibnamefont {Poimanov}}, \bibinfo {author}
  {\bibfnamefont {S.}~\bibnamefont {Sahoo}}, \bibinfo {author} {\bibfnamefont
  {J.}~\bibnamefont {Sinha}}, \bibinfo {author} {\bibfnamefont
  {H.}~\bibnamefont {Yang}}, \bibinfo {author} {\bibfnamefont {M.}~\bibnamefont
  {M\"{u}nzenberg}}, \bibinfo {author} {\bibfnamefont {T.}~\bibnamefont
  {Moriyama}}, \bibinfo {author} {\bibfnamefont {S.}~\bibnamefont {Mizukami}},
  \bibinfo {author} {\bibfnamefont {P.}~\bibnamefont {Landeros}}, \bibinfo
  {author} {\bibfnamefont {R.~A.}\ \bibnamefont {Gallardo}}, \bibinfo {author}
  {\bibfnamefont {G.}~\bibnamefont {Carlotti}}, \bibinfo {author}
  {\bibfnamefont {J.-V.}\ \bibnamefont {Kim}}, \bibinfo {author} {\bibfnamefont
  {R.~L.}\ \bibnamefont {Stamps}}, \bibinfo {author} {\bibfnamefont {R.~E.}\
  \bibnamefont {Camley}}, \bibinfo {author} {\bibfnamefont {B.}~\bibnamefont
  {Rana}}, \bibinfo {author} {\bibfnamefont {Y.}~\bibnamefont {Otani}},
  \bibinfo {author} {\bibfnamefont {W.}~\bibnamefont {Yu}}, \bibinfo {author}
  {\bibfnamefont {T.}~\bibnamefont {Yu}}, \bibinfo {author} {\bibfnamefont
  {G.~E.~W.}\ \bibnamefont {Bauer}}, \bibinfo {author} {\bibfnamefont
  {C.}~\bibnamefont {Back}}, \bibinfo {author} {\bibfnamefont {G.~S.}\
  \bibnamefont {Uhrig}}, \bibinfo {author} {\bibfnamefont {O.~V.}\ \bibnamefont
  {Dobrovolskiy}}, \bibinfo {author} {\bibfnamefont {B.}~\bibnamefont
  {Budinska}}, \bibinfo {author} {\bibfnamefont {H.}~\bibnamefont {Qin}},
  \bibinfo {author} {\bibfnamefont {S.~v.}\ \bibnamefont {Dijken}}, \bibinfo
  {author} {\bibfnamefont {A.~V.}\ \bibnamefont {Chumak}}, \bibinfo {author}
  {\bibfnamefont {A.}~\bibnamefont {Khitun}}, \bibinfo {author} {\bibfnamefont
  {D.~E.}\ \bibnamefont {Nikonov}}, \bibinfo {author} {\bibfnamefont {I.~A.}\
  \bibnamefont {Young}}, \bibinfo {author} {\bibfnamefont {B.~W.}\ \bibnamefont
  {Zingsem}}, \ and\ \bibinfo {author} {\bibfnamefont {M.}~\bibnamefont
  {Winklhofer}},\ }\href {\doibase 10.1088/1361-648X/abec1a} {\bibfield
  {journal} {\bibinfo  {journal} {Journal of Physics: Condensed Matter}\
  }\textbf {\bibinfo {volume} {33}},\ \bibinfo {pages} {413001} (\bibinfo
  {year} {2021})},\ \bibinfo {note} {publisher: IOP Publishing}\BibitemShut
  {NoStop}%
\bibitem [{\citenamefont {R\"{u}ckriegel}\ and\ \citenamefont
  {Duine}(2020)}]{ruckriegel_long-range_2020}%
  \BibitemOpen
  \bibfield  {author} {\bibinfo {author} {\bibfnamefont {A.}~\bibnamefont
  {R\"{u}ckriegel}}\ and\ \bibinfo {author} {\bibfnamefont {R.~A.}\
  \bibnamefont {Duine}},\ }\href {\doibase 10.1103/PhysRevLett.124.117201}
  {\bibfield  {journal} {\bibinfo  {journal} {Physical Review Letters}\
  }\textbf {\bibinfo {volume} {124}},\ \bibinfo {pages} {117201} (\bibinfo
  {year} {2020})},\ \bibinfo {note} {publisher: American Physical
  Society}\BibitemShut {NoStop}%
\bibitem [{\citenamefont {Liu}\ \emph {et~al.}(2019)\citenamefont {Liu},
  \citenamefont {Sun}, \citenamefont {Zhang}, \citenamefont {Groesbeck},
  \citenamefont {Mclaughlin},\ and\ \citenamefont
  {Vardeny}}]{liu_observation_2019}%
  \BibitemOpen
  \bibfield  {author} {\bibinfo {author} {\bibfnamefont {H.}~\bibnamefont
  {Liu}}, \bibinfo {author} {\bibfnamefont {D.}~\bibnamefont {Sun}}, \bibinfo
  {author} {\bibfnamefont {C.}~\bibnamefont {Zhang}}, \bibinfo {author}
  {\bibfnamefont {M.}~\bibnamefont {Groesbeck}}, \bibinfo {author}
  {\bibfnamefont {R.}~\bibnamefont {Mclaughlin}}, \ and\ \bibinfo {author}
  {\bibfnamefont {Z.~V.}\ \bibnamefont {Vardeny}},\ }\href {\doibase
  10.1126/sciadv.aax9144} {\bibfield  {journal} {\bibinfo  {journal} {Science
  Advances}\ }\textbf {\bibinfo {volume} {5}},\ \bibinfo {pages} {eaax9144}
  (\bibinfo {year} {2019})}\BibitemShut {NoStop}%
\bibitem [{\citenamefont {Heinrich}\ \emph {et~al.}(2003)\citenamefont
  {Heinrich}, \citenamefont {Tserkovnyak}, \citenamefont {Woltersdorf},
  \citenamefont {Brataas}, \citenamefont {Urban},\ and\ \citenamefont
  {Bauer}}]{heinrich_dynamic_2003}%
  \BibitemOpen
  \bibfield  {author} {\bibinfo {author} {\bibfnamefont {B.}~\bibnamefont
  {Heinrich}}, \bibinfo {author} {\bibfnamefont {Y.}~\bibnamefont
  {Tserkovnyak}}, \bibinfo {author} {\bibfnamefont {G.}~\bibnamefont
  {Woltersdorf}}, \bibinfo {author} {\bibfnamefont {A.}~\bibnamefont
  {Brataas}}, \bibinfo {author} {\bibfnamefont {R.}~\bibnamefont {Urban}}, \
  and\ \bibinfo {author} {\bibfnamefont {G.~E.~W.}\ \bibnamefont {Bauer}},\
  }\href {\doibase 10.1103/PhysRevLett.90.187601} {\bibfield  {journal}
  {\bibinfo  {journal} {Physical Review Letters}\ }\textbf {\bibinfo {volume}
  {90}},\ \bibinfo {pages} {187601} (\bibinfo {year} {2003})}\BibitemShut
  {NoStop}%
\bibitem [{\citenamefont {Zare~Rameshti}\ \emph {et~al.}(2022)\citenamefont
  {Zare~Rameshti}, \citenamefont {Viola~Kusminskiy}, \citenamefont {Haigh},
  \citenamefont {Usami}, \citenamefont {Lachance-Quirion}, \citenamefont
  {Nakamura}, \citenamefont {Hu}, \citenamefont {Tang}, \citenamefont {Bauer},\
  and\ \citenamefont {Blanter}}]{zare_rameshti_cavity_2022}%
  \BibitemOpen
  \bibfield  {author} {\bibinfo {author} {\bibfnamefont {B.}~\bibnamefont
  {Zare~Rameshti}}, \bibinfo {author} {\bibfnamefont {S.}~\bibnamefont
  {Viola~Kusminskiy}}, \bibinfo {author} {\bibfnamefont {J.~A.}\ \bibnamefont
  {Haigh}}, \bibinfo {author} {\bibfnamefont {K.}~\bibnamefont {Usami}},
  \bibinfo {author} {\bibfnamefont {D.}~\bibnamefont {Lachance-Quirion}},
  \bibinfo {author} {\bibfnamefont {Y.}~\bibnamefont {Nakamura}}, \bibinfo
  {author} {\bibfnamefont {C.-M.}\ \bibnamefont {Hu}}, \bibinfo {author}
  {\bibfnamefont {H.~X.}\ \bibnamefont {Tang}}, \bibinfo {author}
  {\bibfnamefont {G.~E.~W.}\ \bibnamefont {Bauer}}, \ and\ \bibinfo {author}
  {\bibfnamefont {Y.~M.}\ \bibnamefont {Blanter}},\ }\href {\doibase
  10.1016/j.physrep.2022.06.001} {\bibfield  {journal} {\bibinfo  {journal}
  {Physics Reports}\ }\textbf {\bibinfo {volume} {979}},\ \bibinfo {pages} {1}
  (\bibinfo {year} {2022})}\BibitemShut {NoStop}%
\bibitem [{\citenamefont {Bozhko}\ \emph {et~al.}(2020)\citenamefont {Bozhko},
  \citenamefont {Vasyuchka}, \citenamefont {Chumak},\ and\ \citenamefont
  {Serga}}]{bozhko_magnon-phonon_2020}%
  \BibitemOpen
  \bibfield  {author} {\bibinfo {author} {\bibfnamefont {D.~A.}\ \bibnamefont
  {Bozhko}}, \bibinfo {author} {\bibfnamefont {V.~I.}\ \bibnamefont
  {Vasyuchka}}, \bibinfo {author} {\bibfnamefont {A.~V.}\ \bibnamefont
  {Chumak}}, \ and\ \bibinfo {author} {\bibfnamefont {A.~A.}\ \bibnamefont
  {Serga}},\ }\href {\doibase 10.1063/10.0000872} {\bibfield  {journal}
  {\bibinfo  {journal} {Low Temperature Physics}\ }\textbf {\bibinfo {volume}
  {46}},\ \bibinfo {pages} {383} (\bibinfo {year} {2020})},\ \bibinfo {note}
  {publisher: American Institute of Physics}\BibitemShut {NoStop}%
\bibitem [{\citenamefont {Wang}\ and\ \citenamefont
  {Hu}(2020)}]{wang_dissipative_2020}%
  \BibitemOpen
  \bibfield  {author} {\bibinfo {author} {\bibfnamefont {Y.-P.}\ \bibnamefont
  {Wang}}\ and\ \bibinfo {author} {\bibfnamefont {C.-M.}\ \bibnamefont {Hu}},\
  }\href {\doibase 10.1063/1.5144202} {\bibfield  {journal} {\bibinfo
  {journal} {Journal of Applied Physics}\ }\textbf {\bibinfo {volume} {127}},\
  \bibinfo {pages} {130901} (\bibinfo {year} {2020})},\ \bibinfo {note}
  {publisher: American Institute of Physics}\BibitemShut {NoStop}%
\bibitem [{\citenamefont {Harder}\ \emph {et~al.}(2021)\citenamefont {Harder},
  \citenamefont {Yao}, \citenamefont {Gui},\ and\ \citenamefont
  {Hu}}]{harder_coherent_2021}%
  \BibitemOpen
  \bibfield  {author} {\bibinfo {author} {\bibfnamefont {M.}~\bibnamefont
  {Harder}}, \bibinfo {author} {\bibfnamefont {B.~M.}\ \bibnamefont {Yao}},
  \bibinfo {author} {\bibfnamefont {Y.~S.}\ \bibnamefont {Gui}}, \ and\
  \bibinfo {author} {\bibfnamefont {C.-M.}\ \bibnamefont {Hu}},\ }\href
  {\doibase 10.1063/5.0046202} {\bibfield  {journal} {\bibinfo  {journal}
  {Journal of Applied Physics}\ }\textbf {\bibinfo {volume} {129}},\ \bibinfo
  {pages} {201101} (\bibinfo {year} {2021})},\ \bibinfo {note} {publisher:
  American Institute of Physics}\BibitemShut {NoStop}%
\bibitem [{\citenamefont {Zhang}\ \emph {et~al.}(2015)\citenamefont {Zhang},
  \citenamefont {Zou}, \citenamefont {Zhu}, \citenamefont {Marquardt},
  \citenamefont {Jiang},\ and\ \citenamefont {Tang}}]{zhang_magnon_2015}%
  \BibitemOpen
  \bibfield  {author} {\bibinfo {author} {\bibfnamefont {X.}~\bibnamefont
  {Zhang}}, \bibinfo {author} {\bibfnamefont {C.-L.}\ \bibnamefont {Zou}},
  \bibinfo {author} {\bibfnamefont {N.}~\bibnamefont {Zhu}}, \bibinfo {author}
  {\bibfnamefont {F.}~\bibnamefont {Marquardt}}, \bibinfo {author}
  {\bibfnamefont {L.}~\bibnamefont {Jiang}}, \ and\ \bibinfo {author}
  {\bibfnamefont {H.~X.}\ \bibnamefont {Tang}},\ }\href {\doibase
  10.1038/ncomms9914} {\bibfield  {journal} {\bibinfo  {journal} {Nature
  Communications}\ }\textbf {\bibinfo {volume} {6}},\ \bibinfo {pages} {8914}
  (\bibinfo {year} {2015})}\BibitemShut {NoStop}%
\bibitem [{\citenamefont {Lambert}\ \emph {et~al.}(2016)\citenamefont
  {Lambert}, \citenamefont {Haigh}, \citenamefont {Langenfeld}, \citenamefont
  {Doherty},\ and\ \citenamefont {Ferguson}}]{lambert_cavity-mediated_2016}%
  \BibitemOpen
  \bibfield  {author} {\bibinfo {author} {\bibfnamefont {N.~J.}\ \bibnamefont
  {Lambert}}, \bibinfo {author} {\bibfnamefont {J.~A.}\ \bibnamefont {Haigh}},
  \bibinfo {author} {\bibfnamefont {S.}~\bibnamefont {Langenfeld}}, \bibinfo
  {author} {\bibfnamefont {A.~C.}\ \bibnamefont {Doherty}}, \ and\ \bibinfo
  {author} {\bibfnamefont {A.~J.}\ \bibnamefont {Ferguson}},\ }\href {\doibase
  10.1103/PhysRevA.93.021803} {\bibfield  {journal} {\bibinfo  {journal}
  {Physical Review A}\ }\textbf {\bibinfo {volume} {93}},\ \bibinfo {pages}
  {021803} (\bibinfo {year} {2016})}\BibitemShut {NoStop}%
\bibitem [{\citenamefont {Zare~Rameshti}\ and\ \citenamefont
  {Bauer}(2018)}]{zare_rameshti_indirect_2018}%
  \BibitemOpen
  \bibfield  {author} {\bibinfo {author} {\bibfnamefont {B.}~\bibnamefont
  {Zare~Rameshti}}\ and\ \bibinfo {author} {\bibfnamefont {G.~E.~W.}\
  \bibnamefont {Bauer}},\ }\href {\doibase 10.1103/PhysRevB.97.014419}
  {\bibfield  {journal} {\bibinfo  {journal} {Physical Review B}\ }\textbf
  {\bibinfo {volume} {97}},\ \bibinfo {pages} {014419} (\bibinfo {year}
  {2018})}\BibitemShut {NoStop}%
\bibitem [{\citenamefont {Yu}\ \emph {et~al.}(2020)\citenamefont {Yu},
  \citenamefont {Yu},\ and\ \citenamefont {Bauer}}]{yu_circulating_2020}%
  \BibitemOpen
  \bibfield  {author} {\bibinfo {author} {\bibfnamefont {W.}~\bibnamefont
  {Yu}}, \bibinfo {author} {\bibfnamefont {T.}~\bibnamefont {Yu}}, \ and\
  \bibinfo {author} {\bibfnamefont {G.~E.~W.}\ \bibnamefont {Bauer}},\ }\href
  {\doibase 10.1103/PhysRevB.102.064416} {\bibfield  {journal} {\bibinfo
  {journal} {Physical Review B}\ }\textbf {\bibinfo {volume} {102}},\ \bibinfo
  {pages} {064416} (\bibinfo {year} {2020})},\ \bibinfo {note} {publisher:
  American Physical Society}\BibitemShut {NoStop}%
\bibitem [{\citenamefont {Bourhill}\ \emph {et~al.}(2023)\citenamefont
  {Bourhill}, \citenamefont {Yu}, \citenamefont {Vlaminck}, \citenamefont
  {Bauer}, \citenamefont {Ruoso},\ and\ \citenamefont
  {Castel}}]{bourhill_generation_2023}%
  \BibitemOpen
  \bibfield  {author} {\bibinfo {author} {\bibfnamefont {J.}~\bibnamefont
  {Bourhill}}, \bibinfo {author} {\bibfnamefont {W.}~\bibnamefont {Yu}},
  \bibinfo {author} {\bibfnamefont {V.}~\bibnamefont {Vlaminck}}, \bibinfo
  {author} {\bibfnamefont {G.~E.~W.}\ \bibnamefont {Bauer}}, \bibinfo {author}
  {\bibfnamefont {G.}~\bibnamefont {Ruoso}}, \ and\ \bibinfo {author}
  {\bibfnamefont {V.}~\bibnamefont {Castel}},\ }\href {\doibase
  10.1103/PhysRevApplied.19.014030} {\bibfield  {journal} {\bibinfo  {journal}
  {Physical Review Applied}\ }\textbf {\bibinfo {volume} {19}},\ \bibinfo
  {pages} {014030} (\bibinfo {year} {2023})},\ \bibinfo {note} {publisher:
  American Physical Society}\BibitemShut {NoStop}%
\bibitem [{\citenamefont {Berk}\ \emph {et~al.}(2019)\citenamefont {Berk},
  \citenamefont {Jaris}, \citenamefont {Yang}, \citenamefont {Dhuey},
  \citenamefont {Cabrini},\ and\ \citenamefont {Schmidt}}]{berk_strongly_2019}%
  \BibitemOpen
  \bibfield  {author} {\bibinfo {author} {\bibfnamefont {C.}~\bibnamefont
  {Berk}}, \bibinfo {author} {\bibfnamefont {M.}~\bibnamefont {Jaris}},
  \bibinfo {author} {\bibfnamefont {W.}~\bibnamefont {Yang}}, \bibinfo {author}
  {\bibfnamefont {S.}~\bibnamefont {Dhuey}}, \bibinfo {author} {\bibfnamefont
  {S.}~\bibnamefont {Cabrini}}, \ and\ \bibinfo {author} {\bibfnamefont
  {H.}~\bibnamefont {Schmidt}},\ }\href {\doibase 10.1038/s41467-019-10545-x}
  {\bibfield  {journal} {\bibinfo  {journal} {Nature Communications}\ }\textbf
  {\bibinfo {volume} {10}} (\bibinfo {year} {2019}),\
  10.1038/s41467-019-10545-x}\BibitemShut {NoStop}%
\bibitem [{\citenamefont {An}\ \emph {et~al.}(2020)\citenamefont {An},
  \citenamefont {Litvinenko}, \citenamefont {Kohno}, \citenamefont {Fuad},
  \citenamefont {Naletov}, \citenamefont {Vila}, \citenamefont {Ebels},
  \citenamefont {de~Loubens}, \citenamefont {Hurdequint}, \citenamefont
  {Beaulieu}, \citenamefont {Ben~Youssef}, \citenamefont {Vukadinovic},
  \citenamefont {Bauer}, \citenamefont {Slavin}, \citenamefont {Tiberkevich},\
  and\ \citenamefont {Klein}}]{an_coherent_2020}%
  \BibitemOpen
  \bibfield  {author} {\bibinfo {author} {\bibfnamefont {K.}~\bibnamefont
  {An}}, \bibinfo {author} {\bibfnamefont {A.~N.}\ \bibnamefont {Litvinenko}},
  \bibinfo {author} {\bibfnamefont {R.}~\bibnamefont {Kohno}}, \bibinfo
  {author} {\bibfnamefont {A.~A.}\ \bibnamefont {Fuad}}, \bibinfo {author}
  {\bibfnamefont {V.~V.}\ \bibnamefont {Naletov}}, \bibinfo {author}
  {\bibfnamefont {L.}~\bibnamefont {Vila}}, \bibinfo {author} {\bibfnamefont
  {U.}~\bibnamefont {Ebels}}, \bibinfo {author} {\bibfnamefont
  {G.}~\bibnamefont {de~Loubens}}, \bibinfo {author} {\bibfnamefont
  {H.}~\bibnamefont {Hurdequint}}, \bibinfo {author} {\bibfnamefont
  {N.}~\bibnamefont {Beaulieu}}, \bibinfo {author} {\bibfnamefont
  {J.}~\bibnamefont {Ben~Youssef}}, \bibinfo {author} {\bibfnamefont
  {N.}~\bibnamefont {Vukadinovic}}, \bibinfo {author} {\bibfnamefont
  {G.~E.~W.}\ \bibnamefont {Bauer}}, \bibinfo {author} {\bibfnamefont {A.~N.}\
  \bibnamefont {Slavin}}, \bibinfo {author} {\bibfnamefont {V.~S.}\
  \bibnamefont {Tiberkevich}}, \ and\ \bibinfo {author} {\bibfnamefont
  {O.}~\bibnamefont {Klein}},\ }\href {\doibase 10.1103/PhysRevB.101.060407}
  {\bibfield  {journal} {\bibinfo  {journal} {Physical Review B}\ }\textbf
  {\bibinfo {volume} {101}},\ \bibinfo {pages} {060407} (\bibinfo {year}
  {2020})},\ \bibinfo {note} {publisher: American Physical Society}\BibitemShut
  {NoStop}%
\bibitem [{\citenamefont {Li}\ \emph {et~al.}(2021)\citenamefont {Li},
  \citenamefont {Zhao}, \citenamefont {Zhang}, \citenamefont {Hoffmann},\ and\
  \citenamefont {Novosad}}]{li_advances_2021}%
  \BibitemOpen
  \bibfield  {author} {\bibinfo {author} {\bibfnamefont {Y.}~\bibnamefont
  {Li}}, \bibinfo {author} {\bibfnamefont {C.}~\bibnamefont {Zhao}}, \bibinfo
  {author} {\bibfnamefont {W.}~\bibnamefont {Zhang}}, \bibinfo {author}
  {\bibfnamefont {A.}~\bibnamefont {Hoffmann}}, \ and\ \bibinfo {author}
  {\bibfnamefont {V.}~\bibnamefont {Novosad}},\ }\href {\doibase
  10.1063/5.0047054} {\bibfield  {journal} {\bibinfo  {journal} {APL
  Materials}\ }\textbf {\bibinfo {volume} {9}},\ \bibinfo {pages} {060902}
  (\bibinfo {year} {2021})},\ \bibinfo {note} {publisher: AIP Publishing LLCAIP
  Publishing}\BibitemShut {NoStop}%
\bibitem [{\citenamefont {An}\ \emph {et~al.}(2022)\citenamefont {An},
  \citenamefont {Kohno}, \citenamefont {Litvinenko}, \citenamefont {Seeger},
  \citenamefont {Naletov}, \citenamefont {Vila}, \citenamefont {de~Loubens},
  \citenamefont {Ben~Youssef}, \citenamefont {Vukadinovic}, \citenamefont
  {Bauer}, \citenamefont {Slavin}, \citenamefont {Tiberkevich},\ and\
  \citenamefont {Klein}}]{an_bright_2022}%
  \BibitemOpen
  \bibfield  {author} {\bibinfo {author} {\bibfnamefont {K.}~\bibnamefont
  {An}}, \bibinfo {author} {\bibfnamefont {R.}~\bibnamefont {Kohno}}, \bibinfo
  {author} {\bibfnamefont {A.}~\bibnamefont {Litvinenko}}, \bibinfo {author}
  {\bibfnamefont {R.}~\bibnamefont {Seeger}}, \bibinfo {author} {\bibfnamefont
  {V.}~\bibnamefont {Naletov}}, \bibinfo {author} {\bibfnamefont
  {L.}~\bibnamefont {Vila}}, \bibinfo {author} {\bibfnamefont {G.}~\bibnamefont
  {de~Loubens}}, \bibinfo {author} {\bibfnamefont {J.}~\bibnamefont
  {Ben~Youssef}}, \bibinfo {author} {\bibfnamefont {N.}~\bibnamefont
  {Vukadinovic}}, \bibinfo {author} {\bibfnamefont {G.}~\bibnamefont {Bauer}},
  \bibinfo {author} {\bibfnamefont {A.}~\bibnamefont {Slavin}}, \bibinfo
  {author} {\bibfnamefont {V.}~\bibnamefont {Tiberkevich}}, \ and\ \bibinfo
  {author} {\bibfnamefont {O.}~\bibnamefont {Klein}},\ }\href {\doibase
  10.1103/PhysRevX.12.011060} {\bibfield  {journal} {\bibinfo  {journal}
  {Physical Review X}\ }\textbf {\bibinfo {volume} {12}},\ \bibinfo {pages}
  {011060} (\bibinfo {year} {2022})},\ \bibinfo {note} {publisher: American
  Physical Society}\BibitemShut {NoStop}%
\bibitem [{\citenamefont {Harder}\ \emph {et~al.}(2018)\citenamefont {Harder},
  \citenamefont {Yang}, \citenamefont {Yao}, \citenamefont {Yu}, \citenamefont
  {Rao}, \citenamefont {Gui}, \citenamefont {Stamps},\ and\ \citenamefont
  {Hu}}]{harder_level_2018}%
  \BibitemOpen
  \bibfield  {author} {\bibinfo {author} {\bibfnamefont {M.}~\bibnamefont
  {Harder}}, \bibinfo {author} {\bibfnamefont {Y.}~\bibnamefont {Yang}},
  \bibinfo {author} {\bibfnamefont {B.}~\bibnamefont {Yao}}, \bibinfo {author}
  {\bibfnamefont {C.}~\bibnamefont {Yu}}, \bibinfo {author} {\bibfnamefont
  {J.}~\bibnamefont {Rao}}, \bibinfo {author} {\bibfnamefont {Y.}~\bibnamefont
  {Gui}}, \bibinfo {author} {\bibfnamefont {R.}~\bibnamefont {Stamps}}, \ and\
  \bibinfo {author} {\bibfnamefont {C.-M.}\ \bibnamefont {Hu}},\ }\href
  {\doibase 10.1103/PhysRevLett.121.137203} {\bibfield  {journal} {\bibinfo
  {journal} {Physical Review Letters}\ }\textbf {\bibinfo {volume} {121}},\
  \bibinfo {pages} {137203} (\bibinfo {year} {2018})}\BibitemShut {NoStop}%
\bibitem [{\citenamefont {Xu}\ \emph {et~al.}(2019)\citenamefont {Xu},
  \citenamefont {Rao}, \citenamefont {Gui}, \citenamefont {Jin},\ and\
  \citenamefont {Hu}}]{xu_cavity-mediated_2019}%
  \BibitemOpen
  \bibfield  {author} {\bibinfo {author} {\bibfnamefont {P.-C.}\ \bibnamefont
  {Xu}}, \bibinfo {author} {\bibfnamefont {J.~W.}\ \bibnamefont {Rao}},
  \bibinfo {author} {\bibfnamefont {Y.~S.}\ \bibnamefont {Gui}}, \bibinfo
  {author} {\bibfnamefont {X.}~\bibnamefont {Jin}}, \ and\ \bibinfo {author}
  {\bibfnamefont {C.-M.}\ \bibnamefont {Hu}},\ }\href {\doibase
  10.1103/PhysRevB.100.094415} {\bibfield  {journal} {\bibinfo  {journal}
  {Physical Review B}\ }\textbf {\bibinfo {volume} {100}},\ \bibinfo {pages}
  {094415} (\bibinfo {year} {2019})}\BibitemShut {NoStop}%
\bibitem [{\citenamefont {Grigoryan}\ \emph {et~al.}(2018)\citenamefont
  {Grigoryan}, \citenamefont {Shen},\ and\ \citenamefont
  {Xia}}]{grigoryan_synchronized_2018}%
  \BibitemOpen
  \bibfield  {author} {\bibinfo {author} {\bibfnamefont {V.~L.}\ \bibnamefont
  {Grigoryan}}, \bibinfo {author} {\bibfnamefont {K.}~\bibnamefont {Shen}}, \
  and\ \bibinfo {author} {\bibfnamefont {K.}~\bibnamefont {Xia}},\ }\href
  {\doibase 10.1103/PhysRevB.98.024406} {\bibfield  {journal} {\bibinfo
  {journal} {Physical Review B}\ }\textbf {\bibinfo {volume} {98}},\ \bibinfo
  {pages} {024406} (\bibinfo {year} {2018})}\BibitemShut {NoStop}%
\bibitem [{\citenamefont {Grigoryan}\ and\ \citenamefont
  {Xia}(2019)}]{grigoryan_cavity-mediated_2019}%
  \BibitemOpen
  \bibfield  {author} {\bibinfo {author} {\bibfnamefont {V.~L.}\ \bibnamefont
  {Grigoryan}}\ and\ \bibinfo {author} {\bibfnamefont {K.}~\bibnamefont
  {Xia}},\ }\href {\doibase 10.1103/PhysRevB.100.014415} {\bibfield  {journal}
  {\bibinfo  {journal} {Physical Review B}\ }\textbf {\bibinfo {volume}
  {100}},\ \bibinfo {pages} {014415} (\bibinfo {year} {2019})}\BibitemShut
  {NoStop}%
\bibitem [{\citenamefont {Cornelissen}\ \emph {et~al.}(2017)\citenamefont
  {Cornelissen}, \citenamefont {Oyanagi}, \citenamefont {Kikkawa},
  \citenamefont {Qiu}, \citenamefont {Kuschel}, \citenamefont {Bauer},
  \citenamefont {van Wees},\ and\ \citenamefont
  {Saitoh}}]{cornelissen_nonlocal_2017}%
  \BibitemOpen
  \bibfield  {author} {\bibinfo {author} {\bibfnamefont {L.~J.}\ \bibnamefont
  {Cornelissen}}, \bibinfo {author} {\bibfnamefont {K.}~\bibnamefont
  {Oyanagi}}, \bibinfo {author} {\bibfnamefont {T.}~\bibnamefont {Kikkawa}},
  \bibinfo {author} {\bibfnamefont {Z.}~\bibnamefont {Qiu}}, \bibinfo {author}
  {\bibfnamefont {T.}~\bibnamefont {Kuschel}}, \bibinfo {author} {\bibfnamefont
  {G.~E.~W.}\ \bibnamefont {Bauer}}, \bibinfo {author} {\bibfnamefont {B.~J.}\
  \bibnamefont {van Wees}}, \ and\ \bibinfo {author} {\bibfnamefont
  {E.}~\bibnamefont {Saitoh}},\ }\href {\doibase 10.1103/PhysRevB.96.104441}
  {\bibfield  {journal} {\bibinfo  {journal} {Physical Review B}\ }\textbf
  {\bibinfo {volume} {96}},\ \bibinfo {pages} {104441} (\bibinfo {year}
  {2017})},\ \bibinfo {note} {publisher: American Physical Society}\BibitemShut
  {NoStop}%
\bibitem [{\citenamefont {Streib}\ \emph {et~al.}(2018)\citenamefont {Streib},
  \citenamefont {Keshtgar},\ and\ \citenamefont {Bauer}}]{streib_damping_2018}%
  \BibitemOpen
  \bibfield  {author} {\bibinfo {author} {\bibfnamefont {S.}~\bibnamefont
  {Streib}}, \bibinfo {author} {\bibfnamefont {H.}~\bibnamefont {Keshtgar}}, \
  and\ \bibinfo {author} {\bibfnamefont {G.~E.}\ \bibnamefont {Bauer}},\ }\href
  {\doibase 10.1103/PhysRevLett.121.027202} {\bibfield  {journal} {\bibinfo
  {journal} {Physical Review Letters}\ }\textbf {\bibinfo {volume} {121}},\
  \bibinfo {pages} {027202} (\bibinfo {year} {2018})}\BibitemShut {NoStop}%
\bibitem [{\citenamefont {Kittel}(1958)}]{kittel_interaction_1958}%
  \BibitemOpen
  \bibfield  {author} {\bibinfo {author} {\bibfnamefont {C.}~\bibnamefont
  {Kittel}},\ }\href {\doibase 10.1103/PhysRev.110.836} {\bibfield  {journal}
  {\bibinfo  {journal} {Physical Review}\ }\textbf {\bibinfo {volume} {110}},\
  \bibinfo {pages} {836} (\bibinfo {year} {1958})}\BibitemShut {NoStop}%
\bibitem [{Note1()}]{Note1}%
  \BibitemOpen
  \bibinfo {note} {$\protect \mathcal {R}(\theta _0,\phi _0)=\left ( \begin
  {smallmatrix} \cos \theta _0\cos \phi _0 & -\sin \phi _0 & \sin \theta _0\cos
  \phi _0 \\ \cos \theta _0\sin \phi _0 & \cos \phi _0 & \sin \theta _0\sin
  \phi _0 \\ -\sin \theta _0 & 0 & \cos \theta _0 \end {smallmatrix}\right
  )$.}\BibitemShut {Stop}%
\bibitem [{\citenamefont {Dreher}\ \emph {et~al.}(2012)\citenamefont {Dreher},
  \citenamefont {Weiler}, \citenamefont {Pernpeintner}, \citenamefont {Huebl},
  \citenamefont {Gross}, \citenamefont {Brandt},\ and\ \citenamefont
  {Goennenwein}}]{dreher_surface_2012}%
  \BibitemOpen
  \bibfield  {author} {\bibinfo {author} {\bibfnamefont {L.}~\bibnamefont
  {Dreher}}, \bibinfo {author} {\bibfnamefont {M.}~\bibnamefont {Weiler}},
  \bibinfo {author} {\bibfnamefont {M.}~\bibnamefont {Pernpeintner}}, \bibinfo
  {author} {\bibfnamefont {H.}~\bibnamefont {Huebl}}, \bibinfo {author}
  {\bibfnamefont {R.}~\bibnamefont {Gross}}, \bibinfo {author} {\bibfnamefont
  {M.~S.}\ \bibnamefont {Brandt}}, \ and\ \bibinfo {author} {\bibfnamefont
  {S.~T.~B.}\ \bibnamefont {Goennenwein}},\ }\href {\doibase
  10.1103/PhysRevB.86.134415} {\bibfield  {journal} {\bibinfo  {journal}
  {Physical Review B}\ }\textbf {\bibinfo {volume} {86}},\ \bibinfo {pages}
  {134415} (\bibinfo {year} {2012})},\ \bibinfo {note} {publisher: American
  Physical Society}\BibitemShut {NoStop}%
\bibitem [{\citenamefont {Stancil}\ and\ \citenamefont
  {Prabhakar}(2009)}]{stancil_spin_2009}%
  \BibitemOpen
  \bibfield  {author} {\bibinfo {author} {\bibfnamefont {D.~D.}\ \bibnamefont
  {Stancil}}\ and\ \bibinfo {author} {\bibfnamefont {A.}~\bibnamefont
  {Prabhakar}},\ }\href@noop {} {\emph {\bibinfo {title} {Spin {Waves}:
  {Theory} and {Applications}}}}\ (\bibinfo  {publisher} {Springer Science $\&$
  Business Media},\ \bibinfo {year} {2009})\BibitemShut {NoStop}%
\bibitem [{\citenamefont {Rinaldi}\ and\ \citenamefont
  {Turilli}(1985)}]{rinaldi_theory_1985}%
  \BibitemOpen
  \bibfield  {author} {\bibinfo {author} {\bibfnamefont {S.}~\bibnamefont
  {Rinaldi}}\ and\ \bibinfo {author} {\bibfnamefont {G.}~\bibnamefont
  {Turilli}},\ }\href {\doibase 10.1103/PhysRevB.31.3051} {\bibfield  {journal}
  {\bibinfo  {journal} {Physical Review B}\ }\textbf {\bibinfo {volume} {31}},\
  \bibinfo {pages} {3051} (\bibinfo {year} {1985})}\BibitemShut {NoStop}%
\bibitem [{\citenamefont {L\"{u}thi}(2007)}]{luthi_physical_2007}%
  \BibitemOpen
  \bibfield  {author} {\bibinfo {author} {\bibfnamefont {B.}~\bibnamefont
  {L\"{u}thi}},\ }\href@noop {} {\emph {\bibinfo {title} {Physical {Acoustics}
  in the {Solid} {State}}}}\ (\bibinfo  {publisher} {Springer Science \&
  Business Media},\ \bibinfo {year} {2007})\BibitemShut {NoStop}%
\bibitem [{SM()}]{SM}%
  \BibitemOpen
  \href@noop {} {}\bibinfo {note} {See Supplemental Material at
  \url{http://link.aps.org/} for derivation of the generalized Hooke's law in
  the presence of magneto-elastic coupling, calculation of theoretical strength
  of coherent coupling and dynamic exchange coupling and supplemental simulations
  about spectrum transition from strong coupling regime towards weak coupling
  regime.}\BibitemShut {Stop}%
\bibitem [{\citenamefont {Sato}\ \emph {et~al.}(2021)\citenamefont {Sato},
  \citenamefont {Yu}, \citenamefont {Streib},\ and\ \citenamefont
  {Bauer}}]{sato_dynamic_2021}%
  \BibitemOpen
  \bibfield  {author} {\bibinfo {author} {\bibfnamefont {T.}~\bibnamefont
  {Sato}}, \bibinfo {author} {\bibfnamefont {W.}~\bibnamefont {Yu}}, \bibinfo
  {author} {\bibfnamefont {S.}~\bibnamefont {Streib}}, \ and\ \bibinfo {author}
  {\bibfnamefont {G.~E.~W.}\ \bibnamefont {Bauer}},\ }\href {\doibase
  10.1103/PhysRevB.104.014403} {\bibfield  {journal} {\bibinfo  {journal}
  {Physical Review B}\ }\textbf {\bibinfo {volume} {104}},\ \bibinfo {pages}
  {014403} (\bibinfo {year} {2021})},\ \bibinfo {note} {publisher: American
  Physical Society}\BibitemShut {NoStop}%
\bibitem [{com()}]{comsol}%
  \BibitemOpen
  \href {https://www.comsol.com/} {\enquote {\bibinfo {title} {{COMSOL}
  {Multiphysics}{\circledR} v. 5.6. www.comsol.com. {COMSOL AB}, {Stockholm},
  {Sweden}.}}\ }\BibitemShut {NoStop}%
\bibitem [{\citenamefont {Schlitz}\ \emph {et~al.}(2022)\citenamefont
  {Schlitz}, \citenamefont {Siegl}, \citenamefont {Sato}, \citenamefont {Yu},
  \citenamefont {Bauer}, \citenamefont {Huebl},\ and\ \citenamefont
  {Goennenwein}}]{schlitz_magnetization_2022}%
  \BibitemOpen
  \bibfield  {author} {\bibinfo {author} {\bibfnamefont {R.}~\bibnamefont
  {Schlitz}}, \bibinfo {author} {\bibfnamefont {L.}~\bibnamefont {Siegl}},
  \bibinfo {author} {\bibfnamefont {T.}~\bibnamefont {Sato}}, \bibinfo {author}
  {\bibfnamefont {W.}~\bibnamefont {Yu}}, \bibinfo {author} {\bibfnamefont
  {G.~E.~W.}\ \bibnamefont {Bauer}}, \bibinfo {author} {\bibfnamefont
  {H.}~\bibnamefont {Huebl}}, \ and\ \bibinfo {author} {\bibfnamefont
  {S.~T.~B.}\ \bibnamefont {Goennenwein}},\ }\href {\doibase
  10.1103/PhysRevB.106.014407} {\bibfield  {journal} {\bibinfo  {journal}
  {Physical Review B}\ }\textbf {\bibinfo {volume} {106}},\ \bibinfo {pages}
  {014407} (\bibinfo {year} {2022})},\ \bibinfo {note} {publisher: American
  Physical Society}\BibitemShut {NoStop}%
\bibitem [{\citenamefont {Clark}\ and\ \citenamefont
  {Strakna}(1961)}]{clark_elastic_1961}%
  \BibitemOpen
  \bibfield  {author} {\bibinfo {author} {\bibfnamefont {A.~E.}\ \bibnamefont
  {Clark}}\ and\ \bibinfo {author} {\bibfnamefont {R.~E.}\ \bibnamefont
  {Strakna}},\ }\href {\doibase 10.1063/1.1736184} {\bibfield  {journal}
  {\bibinfo  {journal} {Journal of Applied Physics}\ }\textbf {\bibinfo
  {volume} {32}},\ \bibinfo {pages} {1172} (\bibinfo {year}
  {1961})}\BibitemShut {NoStop}%
\bibitem [{\citenamefont {Kleszczewski}\ and\ \citenamefont
  {Bodzenta}(1988)}]{kleszczewski_phononphonon_1988}%
  \BibitemOpen
  \bibfield  {author} {\bibinfo {author} {\bibfnamefont {Z.}~\bibnamefont
  {Kleszczewski}}\ and\ \bibinfo {author} {\bibfnamefont {J.}~\bibnamefont
  {Bodzenta}},\ }\href {\doibase 10.1002/pssb.2221460207} {\bibfield  {journal}
  {\bibinfo  {journal} {physica status solidi (b)}\ }\textbf {\bibinfo {volume}
  {146}},\ \bibinfo {pages} {467} (\bibinfo {year} {1988})},\ \bibinfo {note}
  {\_eprint:
  https://onlinelibrary.wiley.com/doi/pdf/10.1002/pssb.2221460207}\BibitemShut
  {NoStop}%
\bibitem [{\citenamefont {Comstock}(1965)}]{comstock_magnetoelastic_1965}%
  \BibitemOpen
  \bibfield  {author} {\bibinfo {author} {\bibfnamefont {R.~L.}\ \bibnamefont
  {Comstock}},\ }\href {\doibase 10.1109/PROC.1965.4263} {\bibfield  {journal}
  {\bibinfo  {journal} {Proceedings of the IEEE}\ }\textbf {\bibinfo {volume}
  {53}},\ \bibinfo {pages} {1508} (\bibinfo {year} {1965})}\BibitemShut
  {NoStop}%
\bibitem [{Note2()}]{Note2}%
  \BibitemOpen
  \bibinfo {note} {For example, $\protect \bar {\protect \bm {\alpha }}^\prime
  _{(\perp )}=\left ( \begin {smallmatrix} 0 & & \\ & \alpha ^\prime & \\ & &
  \alpha ^\prime \end {smallmatrix}\right )$ for perpendicular magnetization
  along $\protect \hat {\protect \mathbf {x}}$ and $\protect \bar {\protect \bm
  {\alpha }}^\prime _{(\parallel )}=\left ( \begin {smallmatrix} 0 & & \\ &
  \alpha ^\prime & \\ & & 0 \end {smallmatrix}\right )$ for in-plane
  magnetization along $\protect \hat {\protect \mathbf {y}}$}\BibitemShut
  {NoStop}%
\bibitem [{\citenamefont {An}\ \emph {et~al.}(2023)\citenamefont {An},
  \citenamefont {Kim}, \citenamefont {Moon}, \citenamefont {Kohno},
  \citenamefont {Olivetti}, \citenamefont {de~Loubens}, \citenamefont
  {Vukadinovic}, \citenamefont {Youssef}, \citenamefont {Hwang},\ and\
  \citenamefont {Klein}}]{an_optimizing_2023}%
  \BibitemOpen
  \bibfield  {author} {\bibinfo {author} {\bibfnamefont {K.}~\bibnamefont
  {An}}, \bibinfo {author} {\bibfnamefont {C.}~\bibnamefont {Kim}}, \bibinfo
  {author} {\bibfnamefont {K.-W.}\ \bibnamefont {Moon}}, \bibinfo {author}
  {\bibfnamefont {R.}~\bibnamefont {Kohno}}, \bibinfo {author} {\bibfnamefont
  {G.}~\bibnamefont {Olivetti}}, \bibinfo {author} {\bibfnamefont
  {G.}~\bibnamefont {de~Loubens}}, \bibinfo {author} {\bibfnamefont
  {N.}~\bibnamefont {Vukadinovic}}, \bibinfo {author} {\bibfnamefont {J.~B.}\
  \bibnamefont {Youssef}}, \bibinfo {author} {\bibfnamefont {C.}~\bibnamefont
  {Hwang}}, \ and\ \bibinfo {author} {\bibfnamefont {O.}~\bibnamefont
  {Klein}},\ }\href {\doibase 10.48550/arXiv.2302.09936} {\enquote {\bibinfo
  {title} {Optimizing the magnon-phonon cooperativity in planar geometries},}\
  } (\bibinfo {year} {2023}),\ \bibinfo {note} {arXiv:2302.09936 [cond-mat,
  physics:physics]}\BibitemShut {NoStop}%
\bibitem [{\citenamefont {Litvinenko}\ \emph {et~al.}(2021)\citenamefont
  {Litvinenko}, \citenamefont {Khymyn}, \citenamefont {Tyberkevych},
  \citenamefont {Tikhonov}, \citenamefont {Slavin},\ and\ \citenamefont
  {Nikitov}}]{litvinenko_tunable_2021}%
  \BibitemOpen
  \bibfield  {author} {\bibinfo {author} {\bibfnamefont {A.}~\bibnamefont
  {Litvinenko}}, \bibinfo {author} {\bibfnamefont {R.}~\bibnamefont {Khymyn}},
  \bibinfo {author} {\bibfnamefont {V.}~\bibnamefont {Tyberkevych}}, \bibinfo
  {author} {\bibfnamefont {V.}~\bibnamefont {Tikhonov}}, \bibinfo {author}
  {\bibfnamefont {A.}~\bibnamefont {Slavin}}, \ and\ \bibinfo {author}
  {\bibfnamefont {S.}~\bibnamefont {Nikitov}},\ }\href {\doibase
  10.1103/PhysRevApplied.15.034057} {\bibfield  {journal} {\bibinfo  {journal}
  {Physical Review Applied}\ }\textbf {\bibinfo {volume} {15}},\ \bibinfo
  {pages} {034057} (\bibinfo {year} {2021})},\ \bibinfo {note} {publisher:
  American Physical Society}\BibitemShut {NoStop}%
\bibitem [{Note3()}]{Note3}%
  \BibitemOpen
  \bibinfo {note} {$\Omega $ is defined as indirect coherent coupling between
  two magnets meadiated by standing elastic waves, hence $\Omega /\protect
  \sqrt {2}$ indicates direct coupling strength between single magnet and the
  standing wave mode.}\BibitemShut {Stop}%
\bibitem [{\citenamefont {Zhang}\ \emph {et~al.}(2014)\citenamefont {Zhang},
  \citenamefont {Zou}, \citenamefont {Jiang},\ and\ \citenamefont
  {Tang}}]{zhang_strongly_2014}%
  \BibitemOpen
  \bibfield  {author} {\bibinfo {author} {\bibfnamefont {X.}~\bibnamefont
  {Zhang}}, \bibinfo {author} {\bibfnamefont {C.-L.}\ \bibnamefont {Zou}},
  \bibinfo {author} {\bibfnamefont {L.}~\bibnamefont {Jiang}}, \ and\ \bibinfo
  {author} {\bibfnamefont {H.~X.}\ \bibnamefont {Tang}},\ }\href {\doibase
  10.1103/PhysRevLett.113.156401} {\bibfield  {journal} {\bibinfo  {journal}
  {Physical Review Letters}\ }\textbf {\bibinfo {volume} {113}},\ \bibinfo
  {pages} {156401} (\bibinfo {year} {2014})}\BibitemShut {NoStop}%
\bibitem [{\citenamefont {Wang}\ \emph {et~al.}(2019)\citenamefont {Wang},
  \citenamefont {Rao}, \citenamefont {Yang}, \citenamefont {Xu}, \citenamefont
  {Gui}, \citenamefont {Yao}, \citenamefont {You},\ and\ \citenamefont
  {Hu}}]{wang_nonreciprocity_2019}%
  \BibitemOpen
  \bibfield  {author} {\bibinfo {author} {\bibfnamefont {Y.-P.}\ \bibnamefont
  {Wang}}, \bibinfo {author} {\bibfnamefont {J.}~\bibnamefont {Rao}}, \bibinfo
  {author} {\bibfnamefont {Y.}~\bibnamefont {Yang}}, \bibinfo {author}
  {\bibfnamefont {P.-C.}\ \bibnamefont {Xu}}, \bibinfo {author} {\bibfnamefont
  {Y.}~\bibnamefont {Gui}}, \bibinfo {author} {\bibfnamefont {B.}~\bibnamefont
  {Yao}}, \bibinfo {author} {\bibfnamefont {J.}~\bibnamefont {You}}, \ and\
  \bibinfo {author} {\bibfnamefont {C.-M.}\ \bibnamefont {Hu}},\ }\href
  {\doibase 10.1103/PhysRevLett.123.127202} {\bibfield  {journal} {\bibinfo
  {journal} {Physical Review Letters}\ }\textbf {\bibinfo {volume} {123}},\
  \bibinfo {pages} {127202} (\bibinfo {year} {2019})}\BibitemShut {NoStop}%
\bibitem [{\citenamefont {Yu}\ \emph {et~al.}(2019)\citenamefont {Yu},
  \citenamefont {Wang}, \citenamefont {Yuan},\ and\ \citenamefont
  {Xiao}}]{yu_prediction_2019}%
  \BibitemOpen
  \bibfield  {author} {\bibinfo {author} {\bibfnamefont {W.}~\bibnamefont
  {Yu}}, \bibinfo {author} {\bibfnamefont {J.}~\bibnamefont {Wang}}, \bibinfo
  {author} {\bibfnamefont {H.}~\bibnamefont {Yuan}}, \ and\ \bibinfo {author}
  {\bibfnamefont {J.}~\bibnamefont {Xiao}},\ }\href {\doibase
  10.1103/PhysRevLett.123.227201} {\bibfield  {journal} {\bibinfo  {journal}
  {Physical Review Letters}\ }\textbf {\bibinfo {volume} {123}},\ \bibinfo
  {pages} {227201} (\bibinfo {year} {2019})}\BibitemShut {NoStop}%
\bibitem [{\citenamefont {Hu}\ \emph {et~al.}()\citenamefont {Hu},
  \citenamefont {Yu},\ and\ \citenamefont {Wang}}]{hu_auxiliary_nodate}%
  \BibitemOpen
  \bibfield  {author} {\bibinfo {author} {\bibfnamefont {M.-W.}\ \bibnamefont
  {Hu}}, \bibinfo {author} {\bibfnamefont {W.}~\bibnamefont {Yu}}, \ and\
  \bibinfo {author} {\bibfnamefont {Y.-P.}\ \bibnamefont {Wang}},\ }\href
  {\doibase 10.1002/andp.202100534} {\bibfield  {journal} {\bibinfo  {journal}
  {Annalen der Physik}\ }\textbf {\bibinfo {volume} {2100534}},\
  10.1002/andp.202100534}\BibitemShut {NoStop}%
\bibitem [{wan()}]{wang_unpublished_2023}%
  \BibitemOpen
  \href@noop {} {}\bibinfo {note} {J. Wang, and J. Xiao,
  Unpublished.}\BibitemShut {Stop}%
\bibitem [{\citenamefont {Xu}\ \emph {et~al.}(2020)\citenamefont {Xu},
  \citenamefont {Yamamoto}, \citenamefont {Puebla}, \citenamefont {Baumgaertl},
  \citenamefont {Rana}, \citenamefont {Miura}, \citenamefont {Takahashi},
  \citenamefont {Grundler}, \citenamefont {Maekawa},\ and\ \citenamefont
  {Otani}}]{xu_nonreciprocal_2020}%
  \BibitemOpen
  \bibfield  {author} {\bibinfo {author} {\bibfnamefont {M.}~\bibnamefont
  {Xu}}, \bibinfo {author} {\bibfnamefont {K.}~\bibnamefont {Yamamoto}},
  \bibinfo {author} {\bibfnamefont {J.}~\bibnamefont {Puebla}}, \bibinfo
  {author} {\bibfnamefont {K.}~\bibnamefont {Baumgaertl}}, \bibinfo {author}
  {\bibfnamefont {B.}~\bibnamefont {Rana}}, \bibinfo {author} {\bibfnamefont
  {K.}~\bibnamefont {Miura}}, \bibinfo {author} {\bibfnamefont
  {H.}~\bibnamefont {Takahashi}}, \bibinfo {author} {\bibfnamefont
  {D.}~\bibnamefont {Grundler}}, \bibinfo {author} {\bibfnamefont
  {S.}~\bibnamefont {Maekawa}}, \ and\ \bibinfo {author} {\bibfnamefont
  {Y.}~\bibnamefont {Otani}},\ }\href {\doibase 10.1126/sciadv.abb1724}
  {\bibfield  {journal} {\bibinfo  {journal} {Science Advances}\ }\textbf
  {\bibinfo {volume} {6}},\ \bibinfo {pages} {eabb1724} (\bibinfo {year}
  {2020})},\ \bibinfo {note} {publisher: American Association for the
  Advancement of Science Section: Research Article}\BibitemShut {NoStop}%
\bibitem [{\citenamefont {Yamamoto}\ \emph {et~al.}(2020)\citenamefont
  {Yamamoto}, \citenamefont {Yu}, \citenamefont {Yu}, \citenamefont {Puebla},
  \citenamefont {Xu}, \citenamefont {Maekawa},\ and\ \citenamefont
  {Bauer}}]{yamamoto_non-reciprocal_2020}%
  \BibitemOpen
  \bibfield  {author} {\bibinfo {author} {\bibfnamefont {K.}~\bibnamefont
  {Yamamoto}}, \bibinfo {author} {\bibfnamefont {W.}~\bibnamefont {Yu}},
  \bibinfo {author} {\bibfnamefont {T.}~\bibnamefont {Yu}}, \bibinfo {author}
  {\bibfnamefont {J.}~\bibnamefont {Puebla}}, \bibinfo {author} {\bibfnamefont
  {M.}~\bibnamefont {Xu}}, \bibinfo {author} {\bibfnamefont {S.}~\bibnamefont
  {Maekawa}}, \ and\ \bibinfo {author} {\bibfnamefont {G.}~\bibnamefont
  {Bauer}},\ }\href {\doibase 10.7566/JPSJ.89.113702} {\bibfield  {journal}
  {\bibinfo  {journal} {Journal of the Physical Society of Japan}\ }\textbf
  {\bibinfo {volume} {89}},\ \bibinfo {pages} {113702} (\bibinfo {year}
  {2020})},\ \bibinfo {note} {publisher: The Physical Society of
  Japan}\BibitemShut {NoStop}%
\bibitem [{\citenamefont {Yu}(2020)}]{yu_nonreciprocal_2020}%
  \BibitemOpen
  \bibfield  {author} {\bibinfo {author} {\bibfnamefont {T.}~\bibnamefont
  {Yu}},\ }\href {\doibase 10.1103/PhysRevB.102.134417} {\bibfield  {journal}
  {\bibinfo  {journal} {Physical Review B}\ }\textbf {\bibinfo {volume}
  {102}},\ \bibinfo {pages} {134417} (\bibinfo {year} {2020})},\ \bibinfo
  {note} {publisher: American Physical Society}\BibitemShut {NoStop}%
\bibitem [{\citenamefont {Christensen}\ \emph {et~al.}(2016)\citenamefont
  {Christensen}, \citenamefont {Willatzen}, \citenamefont {Velasco},\ and\
  \citenamefont {Lu}}]{christensen_parity-time_2016}%
  \BibitemOpen
  \bibfield  {author} {\bibinfo {author} {\bibfnamefont {J.}~\bibnamefont
  {Christensen}}, \bibinfo {author} {\bibfnamefont {M.}~\bibnamefont
  {Willatzen}}, \bibinfo {author} {\bibfnamefont {V.}~\bibnamefont {Velasco}},
  \ and\ \bibinfo {author} {\bibfnamefont {M.-H.}\ \bibnamefont {Lu}},\ }\href
  {\doibase 10.1103/PhysRevLett.116.207601} {\bibfield  {journal} {\bibinfo
  {journal} {Physical Review Letters}\ }\textbf {\bibinfo {volume} {116}},\
  \bibinfo {pages} {207601} (\bibinfo {year} {2016})},\ \bibinfo {note}
  {publisher: American Physical Society}\BibitemShut {NoStop}%
\bibitem [{\citenamefont {Tserkovnyak}(2020)}]{tserkovnyak_exceptional_2020}%
  \BibitemOpen
  \bibfield  {author} {\bibinfo {author} {\bibfnamefont {Y.}~\bibnamefont
  {Tserkovnyak}},\ }\href {\doibase 10.1103/PhysRevResearch.2.013031}
  {\bibfield  {journal} {\bibinfo  {journal} {Physical Review Research}\
  }\textbf {\bibinfo {volume} {2}},\ \bibinfo {pages} {013031} (\bibinfo {year}
  {2020})}\BibitemShut {NoStop}%
\bibitem [{\citenamefont {Bandyopadhyay}\ \emph {et~al.}(2021)\citenamefont
  {Bandyopadhyay}, \citenamefont {Atulasimha},\ and\ \citenamefont
  {Barman}}]{bandyopadhyay_magnetic_2021}%
  \BibitemOpen
  \bibfield  {author} {\bibinfo {author} {\bibfnamefont {S.}~\bibnamefont
  {Bandyopadhyay}}, \bibinfo {author} {\bibfnamefont {J.}~\bibnamefont
  {Atulasimha}}, \ and\ \bibinfo {author} {\bibfnamefont {A.}~\bibnamefont
  {Barman}},\ }\href {\doibase 10.1063/5.0062993} {\bibfield  {journal}
  {\bibinfo  {journal} {Applied Physics Reviews}\ }\textbf {\bibinfo {volume}
  {8}},\ \bibinfo {pages} {041323} (\bibinfo {year} {2021})},\ \bibinfo {note}
  {publisher: American Institute of Physics}\BibitemShut {NoStop}%
\bibitem [{\citenamefont {Zhang}\ \emph {et~al.}(2022)\citenamefont {Zhang},
  \citenamefont {Yang}, \citenamefont {Sheng},\ and\ \citenamefont
  {Wu}}]{zhang_dissipative_2022}%
  \BibitemOpen
  \bibfield  {author} {\bibinfo {author} {\bibfnamefont {Q.}~\bibnamefont
  {Zhang}}, \bibinfo {author} {\bibfnamefont {C.}~\bibnamefont {Yang}},
  \bibinfo {author} {\bibfnamefont {J.}~\bibnamefont {Sheng}}, \ and\ \bibinfo
  {author} {\bibfnamefont {H.}~\bibnamefont {Wu}},\ }\href {\doibase
  10.1073/pnas.2207543119} {\bibfield  {journal} {\bibinfo  {journal}
  {Proceedings of the National Academy of Sciences}\ }\textbf {\bibinfo
  {volume} {119}},\ \bibinfo {pages} {e2207543119} (\bibinfo {year} {2022})},\
  \bibinfo {note} {publisher: Proceedings of the National Academy of
  Sciences}\BibitemShut {NoStop}%
\end{thebibliography}

%

\clearpage


\section*{Supplementary material}

# Supplemental Materials: Dynamic Exchange Coupling between Magnets Mediated by Attenuating Elastic Waves

Weichao Yu (余伟超)[1,2,*]

[1]*State Key Laboratory of Surface Physics and Institute for Nanoelectronic Devices and Quantum Computing, Fudan University, Shanghai 200433, China*
[2]*Zhangjiang Fudan International Innovation Center, Fudan University, Shanghai 201210, China*

**CONTENTS**

* wcyu@fudan.edu.cn

# I. GENERALIZED HOOKE'S LAW IN THE PRESENCE OF MAGNETO-ELASTIC COUPLING

The numerical approach applied in the main text is a new method to perform simulation in hybrid magnon-phonon systems, whose concept is to treat the dynamics and linear response of magnetic system as an effective elastic stiffness tensor. Armed with this method and with the help of finite-element method, one is able to simulate phononic excitations in the presence of magnets in arbitrary shapes and positioned in arbitrary configurations.

## A. Effective field induced by magneto-elastic coupling

We first define the vector of magnetization as $\mathbf{M} = M_s\mathbf{m}$, with $M_s$ the saturated magnetization and the uniform magnetic order $\mathbf{m} = (m_x, m_y, m_z)$. In solids, the elastic degree of freedom is captured by the strain tensor, which is defined as $\varepsilon_{ij} = \frac{1}{2}\left(\frac{\partial u_i}{\partial x_j} + \frac{\partial u_j}{\partial x_i}\right)$ with $\mathbf{u} = \{u_x, u_y, u_z\}$ the lattice displacement away from its equilibrium position. The energy density in the presence of magneto-elastic coupling for cubic system is given by [1]

$$
\begin{aligned}
\mathcal{U}_{\text{me}} &= b_1\sum_i m_i^2\varepsilon_{ii} + b_2\sum_{i\neq j} m_i m_j\varepsilon_{ij} \\
&= b_1 m_x^2\varepsilon_{xx} + b_1 m_y^2\varepsilon_{yy} + b_1 m_z^2\varepsilon_{zz} + 2b_2 m_x m_y\varepsilon_{xy} + 2b_2 m_x m_z\varepsilon_{xz} + 2b_2 m_y m_z\varepsilon_{yz},
\end{aligned}
\tag{S1}
$$

where $b_1$ and $b_2$ are phenomenological coefficients for specific materials.

The transformation between global coordinate $(x, y, z)$ to local coordinate $(1, 2, 3)$ is given by [2]

$$
\begin{pmatrix} m_x \\ m_y \\ m_z \end{pmatrix} = \begin{pmatrix} \cos\theta_0\cos\phi_0 & -\sin\phi_0 & \sin\theta_0\cos\phi_0 \\ \cos\theta_0\sin\phi_0 & \cos\phi_0 & \sin\theta_0\sin\phi_0 \\ -\sin\theta_0 & 0 & \cos\theta_0 \end{pmatrix} \begin{pmatrix} m_1 \\ m_2 \\ m_3 \end{pmatrix},
\tag{S2}
$$

with $\theta_0$ and $\phi_0$ the polar and azimuthal angle with respect to equilibrium magnetization, which is expressed in spherical coordinate as $\mathbf{m}_0 = (\sin\theta_0\cos\phi_0, \sin\theta_0\sin\phi_0, \cos\theta_0)$. In local coordinate with $\mathbf{m} = (m_1, m_2, m_3)$, the dynamics of magnetization is affected by the effective field, which is given by

$$
\delta\mathbf{h} = -\frac{1}{\mu_0 M_s}\frac{\partial\mathcal{U}_{\text{me}}}{\partial(\boldsymbol{\delta}\mathbf{m})},
\tag{S3}
$$

where $\delta\mathbf{h} = (h_1, h_2, h_3)$ also in local coordinate.

Assuming that the magnetization is perturbed around its equilibrium, the magnetization can be decomposed into two parts with $m_3 \simeq |\mathbf{m}_0| = 1$ and

$$
\mathbf{m} \simeq \mathbf{m}_0 + \delta\mathbf{m}e^{i\omega t} = \begin{pmatrix} 0 \\ 0 \\ 1 \end{pmatrix} + \begin{pmatrix} m_1 \\ m_2 \\ 0 \end{pmatrix} e^{i\omega t}.
\tag{S4}
$$

Combining Eq.(S1), Eq.(S2) and Eq.(S3) and neglecting high-order terms, we obtain effective field (dynamic part) induced by magneto-elastic coupling in transverse direction [2]

$$
\begin{aligned}
\mu_0 M_s h_1 &= -2b_1\varepsilon_{xx}\sin\theta_0\cos\theta_0\cos^2\phi_0 + 2b_1\varepsilon_{zz}\sin\theta_0\cos\theta_0 - 2b_1\varepsilon_{yy}\sin\theta_0\cos\theta_0\sin^2\phi_0 \\
&\quad - 4b_2\varepsilon_{xy}\sin\theta_0\cos\theta_0\sin\phi_0\cos\phi_0 - 2b_2\varepsilon_{xz}\cos 2\theta_0\cos\phi_0 - 2b_2\varepsilon_{yz}\cos 2\theta_0\sin\phi_0,
\end{aligned}
\tag{S5a}
$$

$$
\begin{aligned}
\mu_0 M_s h_2 &= -b_1\varepsilon_{yy}\sin\theta\sin 2\phi_0 + b_1\varepsilon_{xx}\sin\theta_0\sin 2\phi_0 \\
&\quad - 2b_2\varepsilon_{xy}\sin\theta_0\cos 2\phi_0 + 2b_2\varepsilon_{xz}\cos\theta_0\sin\phi_0 - 2b_2\varepsilon_{yz}\cos\theta_0\cos\phi_0.
\end{aligned}
\tag{S5b}
$$

There is no dynamics in longitudinal direction since $h_3 = 0$ under linear approximation.

## B. Linear response of magnetization under elastic perturbation

In this part, we will derive the response of magnet under the perturbation of lattice deformation due to magneto-elastic coupling, which is characterized by the Polder susceptibility tensor.

Start from the time-dependent Landau-Lifshitz-Gilbert equation without torque terms

$$\frac{\partial \mathbf{M}}{\partial t} = -\gamma \mathbf{M} \times \mathbf{H}_{\text{eff}} + \frac{\alpha}{M_{\text{s}}} \mathbf{M} \times \frac{\partial \mathbf{M}}{\partial t}, \tag{S6}$$

where $\gamma$ is gyromagnetic ratio and $\alpha$ is Gilbert damping coefficient. We consider the simplest case where only external field presents with $\mathbf{H}_{\text{eff}} = \mathbf{H}$. Assuming that the magnetic moment $\mathbf{m}$ is perturbed around its equilibrium magnetization $\mathbf{M}_0$ with certain angular frequency $\omega$, it can be decomposed as $\mathbf{M} = M_{\text{s}}\mathbf{m}_0 + M_{\text{s}}\delta\mathbf{m}e^{i\omega t}$, with $\delta\mathbf{m} = (m_1, m_2, 0)$ and $\mathbf{m}_0 = (0, 0, 1)$ in local coordinate. The effective field can be decomposed in the same way as $\mathbf{H} = \mathbf{H}_0 + \delta\mathbf{H}e^{i\omega t}$, with $\mathbf{H}_0 = (0, 0, H_0)$ and $\delta\mathbf{H} = (H_1, H_2, 0)$. Plugging back to Eq.(S6) and eliminating static components and common factor $e^{i\omega t}$, we obtain the linearized LLG equation

$$i\omega\delta\mathbf{m} = -\gamma\mathbf{M}_0 \times \delta\mathbf{H} - \gamma\delta\mathbf{m} \times \mathbf{H}_0 + i\omega\alpha\mathbf{m}_0 \times \delta\mathbf{m}, \tag{S7}$$

which can be rearranged as

$$\begin{pmatrix} H_1 \\ H_2 \end{pmatrix} = \frac{M_{\text{s}}}{\omega_{\text{M}}} \begin{pmatrix} \omega_0 - i\omega\alpha & i\omega \\ -i\omega & \omega_0 - i\omega\alpha \end{pmatrix} \begin{pmatrix} m_1 \\ m_2 \end{pmatrix}, \tag{S8}$$

with the definition $\omega_0 = \gamma H_0$ and $\omega_{\text{M}} = \gamma M_{\text{s}}$. Eq.(S8) can be further rearranged in the form

$$M_{\text{s}}\delta\mathbf{m} = \bar{\boldsymbol{\chi}}\delta\mathbf{H}, \tag{S9}$$

with $\bar{\boldsymbol{\chi}}$ the *Polder susceptibility tensor* [3] whose components are

$$\chi_{11} = \chi_{22} = \frac{\omega_{\text{M}}(\omega_0 - i\omega\alpha)}{(\omega_0 - i\omega\alpha)^2 - \omega^2}, \tag{S10a}$$

$$\chi_{12} = -\chi_{21} = \frac{-i\omega\omega_{\text{M}}}{(\omega_0 - i\omega\alpha)^2 - \omega^2}. \tag{S10b}$$

With the susceptibility tensor Eq.(S10), we are able to describe the response of magnetization under external perturbation. However, the perturbation is not restricted to external microwave fields, which can be induced by strain in the form of effective field as we have already derived in Eq.(S3). Hence we can rewrite Eq.(S9) as

$$\begin{aligned}
\begin{pmatrix} m_1 \\ m_2 \end{pmatrix} &= \frac{1}{M_{\text{s}}} \begin{pmatrix} \chi_{11} & \chi_{12} \\ \chi_{21} & \chi_{22} \end{pmatrix} \begin{pmatrix} H_1 + h_1 \\ H_2 + h_2 \end{pmatrix} \\
&= \begin{pmatrix} \widetilde{\chi}_{11} & \widetilde{\chi}_{12} \\ \widetilde{\chi}_{21} & \widetilde{\chi}_{22} \end{pmatrix} \begin{pmatrix} \widetilde{H}_1 + \widetilde{h}_1 \\ \widetilde{H}_2 + \widetilde{h}_2 \end{pmatrix}
\end{aligned} \tag{S11}$$

where $H_{1,2}$ are external (dynamical) fields and $h_{1,2}$ are effective fields (Eq.(S3)) induced by strain in the presence of magneto-elastic coupling. In the second line, we perform the transformation that $\widetilde{\chi} = \chi/(\mu_0 M_{\text{s}}^2)$, $\delta\widetilde{\mathbf{H}} = \mu_0 M_{\text{s}}\delta\mathbf{H}$ and $\delta\widetilde{\mathbf{h}} = \mu_0 M_{\text{s}}\delta\mathbf{h}$, so that $\delta\widetilde{\mathbf{H}}$ and $\delta\widetilde{\mathbf{h}}$ are *effective stress*, and $\widetilde{\chi}$ is the susceptibility tensor relating effective stress and dimensionless magnetization excitation. Plugging Eq.(S11) back to Eq.(S1), we obtain an explicit form of energy density in terms of static magnetization $\theta_0$, $\phi_0$ and strain $\varepsilon_{ij}$, which includes the backaction of magnetization excitation in the presence of dynamic strain. Since the strain $\varepsilon_{ij}$ is spatially dependent, the magnetization excitation is inhomogeneous even though the static magnetization is uniform, leading to larger strength for coherent coupling and dynamic exchange coupling when compared to simplified theoretical model.

## C. Effective stiffness tensor in the presence of magneto-elastic coupling

According to theory of elasticity [2, 4], the magneto-elastic energy density can be expressed with Einstein summation convention as

$$\mathcal{U}_{\text{me}} = \frac{1}{2} C^{\text{me}}_{ijkl} \varepsilon_{ij} \varepsilon_{kl}, \tag{S12}$$

with $C^{\text{me}}_{ijkl}$ the fourth-rank stiffness tensor (or elasticity tensor) contributed from magneto-elastic coupling. We can further obtain the corresponding stress tensor

$$\sigma^{\text{me}}_{ij} = \frac{\partial \mathcal{U}_{\text{me}}}{\partial \varepsilon_{ij}} = C^{\text{me}}_{ijkl} \varepsilon_{kl}, \tag{S13}$$

which is essentially the generalized Hooke's law. Due to their inherent symmetries, the number of degrees of freedom can be further reduced, so that $C^{\text{me}}_{ijkl} \to C^{\text{me}}_{ij}$ and $\sigma^{\text{me}}_{ij}, \varepsilon^{\text{me}}_{ij} \to \sigma^{\text{me}}_i, \varepsilon^{\text{me}}_i$. Here we follow the convention of *Voigt notation* by defining

$$\sigma_i = \begin{pmatrix} \sigma_1 \\ \sigma_2 \\ \sigma_3 \\ \sigma_4 \\ \sigma_5 \\ \sigma_6 \end{pmatrix} = \begin{pmatrix} \sigma_{xx} \\ \sigma_{yy} \\ \sigma_{zz} \\ \sigma_{yz} \\ \sigma_{xz} \\ \sigma_{xy} \end{pmatrix}, \quad \varepsilon_i = \begin{pmatrix} \varepsilon_1 \\ \varepsilon_2 \\ \varepsilon_3 \\ \varepsilon_4 \\ \varepsilon_5 \\ \varepsilon_6 \end{pmatrix} = \begin{pmatrix} \varepsilon_{xx} \\ \varepsilon_{yy} \\ \varepsilon_{zz} \\ 2\varepsilon_{yz} \\ 2\varepsilon_{xz} \\ 2\varepsilon_{xy} \end{pmatrix}. \tag{S14}$$

Now we move on to calculate all six components of the stress vector $\sigma^{\text{me}}_i$ according to $\sigma^{\text{me}}_i = \partial \mathcal{U}_{\text{me}}/\partial \varepsilon_i$ with combination of effective fields induced by magneto-elastic coupling Eq.(S3) and magnetization response under the effective fields as well external fields Eq.(S11). By neglecting high-order terms such as $\sim \varepsilon^2$ and $\sim \widetilde{\chi}^2$, we obtain

$$\sigma^{\text{me}}_{xx} =$$
$$\left[ b_1 \sin^2\theta_0 \cos^2\phi_0 + 2b_1 \widetilde{H}_2 \sin\theta_0 \left( \widetilde{\chi}_{12} \cos\theta_0 \cos^2\phi_0 - \widetilde{\chi}_{22} \sin\phi_0 \cos\phi_0 \right) + 2b_1 \widetilde{H}_1 \sin\theta_0 \cos\phi_0 \left( \widetilde{\chi}_{11} \cos\theta_0 \cos\phi_0 - \widetilde{\chi}_{21} \sin\phi_0 \right) \right]$$
$$+ \varepsilon_{xx} \left[ -2b_1^2 \sin\theta_0 \left( \widetilde{\chi}_{22} \sin\theta_0 \sin^2(2\phi_0) - 2\widetilde{\chi}_{12} \sin(2\theta_0) \sin\phi_0 \cos^3\phi_0 - 2\widetilde{\chi}_{21} \sin(2\theta_0) \sin\phi_0 \cos^3\phi_0 + 4\widetilde{\chi}_{11} \sin\theta_0 \cos^2\theta_0 \cos^4\phi_0 \right) \right]$$
$$+ \varepsilon_{yy} \left[ -\frac{1}{2} b_1^2 \sin\theta_0 \left( -4\widetilde{\chi}_{22} \sin\theta_0 \sin^2(2\phi_0) + \widetilde{\chi}_{12} \sin(2\theta_0) \sin(4\phi_0) + \widetilde{\chi}_{21} \sin(2\theta_0) \sin(4\phi_0) + 4\widetilde{\chi}_{11} \sin\theta_0 \cos^2\theta_0 \sin^2(2\phi_0) \right) \right]$$
$$+ \varepsilon_{zz} \left[ -4b_1^2 \sin^2\theta_0 \cos\theta_0 \cos\phi_0 \left( (\widetilde{\chi}_{12} + \widetilde{\chi}_{21}) \sin\phi_0 - 2\widetilde{\chi}_{11} \cos\theta_0 \cos\phi_0 \right) \right]$$
$$+ \varepsilon_{yz} \left[ b_1 \sin\theta_0 \left( -2b_2 \widetilde{\chi}_{12} \left( \sin^2\theta_0 \sin\phi_0 \sin(2\phi_0) + \cos^2\theta_0 (\cos\phi_0 + \cos(3\phi_0)) \right) - 8b_2 \widetilde{\chi}_{11} \cos\theta_0 \cos(2\theta_0) \sin\phi_0 \cos^2\phi_0 \right. \right.$$
$$\left. \left. + 8b_2 \widetilde{\chi}_{22} \cos\theta_0 \sin\phi_0 \cos^2\phi_0 + b_2 \widetilde{\chi}_{21} \left( 4\cos(2\theta_0) \sin^2\phi_0 \cos\phi_0 - 4\cos^2\theta_0 \cos^3\phi_0 \right) \right) \right]$$
$$+ \varepsilon_{xz} \left[ b_1 \sin\theta_0 \left( -8b_2 \widetilde{\chi}_{11} \cos\theta_0 \cos(2\theta_0) \cos^3\phi_0 + 2b_2 \widetilde{\chi}_{12} (3\cos(2\theta_0) + 1) \sin\phi_0 \cos^2\phi_0 + 2b_2 \widetilde{\chi}_{21} (3\cos(2\theta_0) + 1) \sin\phi_0 \cos^2\phi_0 \right. \right.$$
$$\left. \left. - 8b_2 \widetilde{\chi}_{22} \cos\theta_0 \sin^2\phi_0 \cos\phi_0 \right) \right]$$
$$+ \varepsilon_{xy} \left[ b_1 \sin\theta_0 \left( 2b_2 \widetilde{\chi}_{22} \sin\theta_0 \sin(4\phi_0) + 2b_2 \widetilde{\chi}_{12} \sin(2\theta_0)(1 - 2\cos(2\phi_0)) \cos^2\phi_0 + 2b_2 \widetilde{\chi}_{21} \sin(2\theta_0)(1 - 2\cos(2\phi_0)) \cos^2\phi_0 \right. \right.$$
$$\left. \left. - 16b_2 \widetilde{\chi}_{11} \sin\theta_0 \cos^2\theta_0 \sin\phi_0 \cos^3\phi_0 \right) \right], \tag{S15}$$

$$\sigma_{yy}^{\mathrm{me}} =$$
$$\left[ b_1 \sin^2\theta_0 \sin^2\phi_0 + b_1 \widetilde{H}_1 \sin\theta_0 \left( 2\widetilde{\chi}_{11} \cos\theta_0 \sin^2\phi_0 + \widetilde{\chi}_{21} \sin(2\phi_0) \right) + b_1 \widetilde{H}_2 \sin\theta_0 \left( 2\widetilde{\chi}_{12} \cos\theta_0 \sin^2\phi_0 + \widetilde{\chi}_{22} \sin(2\phi_0) \right) \right]$$
$$+ \varepsilon_{xx} \left[ -\frac{1}{2} b_1^2 \sin\theta_0 \left( -4\widetilde{\chi}_{22} \sin\theta_0 \sin^2(2\phi_0) + \widetilde{\chi}_{12} \sin(2\theta_0) \sin(4\phi_0) + \widetilde{\chi}_{21} \sin(2\theta_0) \sin(4\phi_0) + 4\widetilde{\chi}_{11} \sin\theta_0 \cos^2\theta_0 \sin^2(2\phi_0) \right) \right]$$
$$+ \varepsilon_{yy} \left[ -2b_1^2 \sin\theta_0 \left( \widetilde{\chi}_{22} \sin\theta_0 \sin^2(2\phi_0) + 4\widetilde{\chi}_{11} \sin\theta_0 \cos^2\theta_0 \sin^4\phi_0 + 2\widetilde{\chi}_{12} \sin(2\theta_0) \sin^3\phi_0 \cos\phi_0 + 2\widetilde{\chi}_{21} \sin(2\theta_0) \sin^3\phi_0 \cos\phi_0 \right) \right]$$
$$+ \varepsilon_{zz} \left[ 2b_1^2 \sin\theta_0 \sin(2\theta_0) \sin\phi_0 \left( 2\widetilde{\chi}_{11} \cos\theta_0 \sin\phi_0 + (\widetilde{\chi}_{12} + \widetilde{\chi}_{21}) \cos\phi_0 \right) \right]$$
$$+ \varepsilon_{yz} \left[ b_1 \sin\theta_0 \left( -8b_2\widetilde{\chi}_{22} \cos\theta_0 \cos^2\phi_0 - 8b_2\widetilde{\chi}_{11} \cos\theta_0 \cos(2\theta_0) \sin^3\phi_0 - 2b_2\widetilde{\chi}_{12}(3\cos(2\theta_0) + 1)\sin^2\phi_0 \cos\phi_0 \right. \right.$$
$$\left. \left. -2b_2\widetilde{\chi}_{21}(3\cos(2\theta_0) + 1)\sin^2\phi_0 \cos\phi_0 \right) \right]$$
$$+ \varepsilon_{xx} \left[ b_1 \sin\theta_0 \left( b_2\widetilde{\chi}_{21} \left( 4\cos^2\theta_0 \sin^3\phi_0 - 4\cos(2\theta_0)\sin\phi_0 \cos^2\phi_0 \right) + b_2\widetilde{\chi}_{12} \left( 4\sin^2\theta_0 \sin\phi_0 \cos^2\phi_0 + 2\cos^2\theta_0(\sin\phi_0 - \sin(3\phi_0)) \right) \right) \right.$$
$$\left. -8b_2\widetilde{\chi}_{11} \cos\theta_0 \cos(2\theta_0)\sin^2\phi_0 \cos\phi_0 + 8b_2\widetilde{\chi}_{22} \cos\theta_0 \sin^2\phi_0 \cos\phi_0 \right) \right]$$
$$+ \varepsilon_{xy} \left[ b_1 \sin\theta_0 \left( -2b_2\widetilde{\chi}_{22} \sin\theta_0 \sin(4\phi_0) - 16b_2\widetilde{\chi}_{11} \sin\theta_0 \cos^2\theta_0 \sin^3\phi_0 \cos\phi_0 - 2b_2\widetilde{\chi}_{12} \sin(2\theta_0)\sin^2\phi_0 (2\cos(2\phi_0) + 1) \right. \right.$$
$$\left. \left. -2b_2\widetilde{\chi}_{21} \sin(2\theta_0)\sin^2\phi_0 (2\cos(2\phi_0) + 1) \right) \right] ,$$

$$\text{(S16)}$$

$$\sigma_{zz}^{\mathrm{me}} =$$
$$\left[ b_1 \cos^2\theta_0 - 2b_1 \widetilde{H}_1 \widetilde{\chi}_{11} \sin\theta_0 \cos\theta_0 - 2b_1 \widetilde{H}_2 \widetilde{\chi}_{12} \sin\theta_0 \cos\theta_0 \right]$$
$$+ \varepsilon_{xx} \left[ 2b_1^2 \sin^2\theta_0 \cos\theta_0 \left( 4\widetilde{\chi}_{11} \cos\theta_0 \cos^2\phi_0 - (\widetilde{\chi}_{12} + \widetilde{\chi}_{21})\sin(2\phi_0) \right) \right]$$
$$+ \varepsilon_{yy} \left[ 2b_1^2 \sin^2\theta_0 \cos\theta_0 \left( 4\widetilde{\chi}_{11} \cos\theta_0 \sin^2\phi_0 + (\widetilde{\chi}_{12} + \widetilde{\chi}_{21})\sin(2\phi_0) \right) \right]$$
$$+ \varepsilon_{zz} \left[ -8b_1^2 \widetilde{\chi}_{11} \sin^2\theta_0 \cos^2\theta_0 \right]$$
$$+ \varepsilon_{yz} \left[ b_1 \cos\theta_0 \left( 8b_2\widetilde{\chi}_{11} \sin\theta_0 \cos(2\theta_0) \sin\phi_0 + 4b_2\widetilde{\chi}_{12} \sin\theta_0 \cos\theta_0 \cos\phi_0 + 4b_2\widetilde{\chi}_{21} \sin\theta_0 \cos\theta_0 \cos\phi_0 \right) \right]$$
$$+ \varepsilon_{xx} \left[ b_1 \cos\theta_0 \left( 8b_2\widetilde{\chi}_{11} \sin\theta_0 \cos(2\theta_0) \cos\phi_0 - 4b_2\widetilde{\chi}_{12} \sin\theta_0 \cos\theta_0 \sin\phi_0 - 4b_2\widetilde{\chi}_{21} \sin\theta_0 \cos\theta_0 \sin\phi_0 \right) \right]$$
$$+ \varepsilon_{xy} \left[ b_1 \cos\theta_0 \left( 8b_2\widetilde{\chi}_{11} \sin^2\theta_0 \cos\theta_0 \sin(2\phi_0) + 4b_2\widetilde{\chi}_{12} \sin^2\theta_0 \cos(2\phi_0) + 4b_2\widetilde{\chi}_{21} \sin^2\theta_0 \cos(2\phi_0) \right) \right] ,$$

$$\text{(S17)}$$

$$\sigma_{yz}^{\mathrm{me}} =$$
$$\left[ b_2 \sin\theta_0 \cos\theta_0 \sin\phi_0 + b_2 \widetilde{H}_1 \left( \widetilde{\chi}_{21} \cos\theta_0 \cos\phi_0 + \widetilde{\chi}_{11} \cos(2\theta_0)\sin\phi_0 \right) + b_2 \widetilde{H}_2 \left( \widetilde{\chi}_{22} \cos\theta_0 \cos\phi_0 + \widetilde{\chi}_{12} \cos(2\theta_0)\sin\phi_0 \right) \right]$$
$$+ \varepsilon_{xx} \left[ b_1 b_2 \left( \widetilde{\chi}_{11} \sin(4\theta_0)\sin\phi_0 \left( -\cos^2\phi_0 \right) - \sin\theta_0 \left( \widetilde{\chi}_{12} \left( \sin^2\theta_0 \sin\phi_0 \sin(2\phi_0) + \cos^2\theta_0 (\cos\phi_0 + \cos(3\phi_0)) \right) \right. \right. \right.$$
$$\left. \left. \left. -4\widetilde{\chi}_{22} \cos\theta_0 \sin\phi_0 \cos^2\phi_0 + \widetilde{\chi}_{21} \left( 2\cos^2\theta_0 \cos^3\phi_0 - 2\cos(2\theta_0)\sin^2\phi_0 \cos\phi_0 \right) \right) \right) \right]$$
$$+ \varepsilon_{yy} \left[ -b_1 b_2 \sin\phi_0 \left( \widetilde{\chi}_{11} \sin(4\theta_0)\sin^2\phi_0 + \sin\theta_0 \cos\phi_0 \left( 4\widetilde{\chi}_{22} \cos\theta_0 \cos\phi_0 + \widetilde{\chi}_{12}(3\cos(2\theta_0) + 1)\sin\phi_0 + \widetilde{\chi}_{21}(3\cos(2\theta_0) + 1)\sin\phi_0 \right) \right) \right]$$
$$+ \varepsilon_{zz} \left[ b_1 b_2 \left( \widetilde{\chi}_{11} \sin(4\theta_0)\sin\phi_0 + 2\left( \widetilde{\chi}_{12} + \widetilde{\chi}_{21} \right) \sin\theta_0 \cos^2\theta_0 \cos\phi_0 \right) \right]$$
$$+ \varepsilon_{yz} \left[ b_2 \left( -4b_2\widetilde{\chi}_{22} \cos^2\theta_0 \cos^2\phi_0 - 4b_2\widetilde{\chi}_{11} \cos^2(2\theta_0)\sin^2\phi_0 - 4b_2\widetilde{\chi}_{21} \cos\theta_0 \cos(2\theta_0)\sin\phi_0 \cos\phi_0 - 2b_2\widetilde{\chi}_{12} \cos\theta_0 \cos(2\theta_0)\sin(2\phi_0) \right) \right]$$
$$+ \varepsilon_{xz} \left[ b_2 \left( -2b_2\widetilde{\chi}_{12} \cos(2\theta_0)\cos\theta_0 \cos(2\phi_0) + b_2\widetilde{\chi}_{21}(-\cos\theta_0 - \cos(3\theta_0))\cos(2\phi_0) + 4b_2\widetilde{\chi}_{22} \cos^2\theta_0 \sin\phi_0 \cos\phi_0 \right. \right.$$
$$\left. \left. -2b_2\widetilde{\chi}_{11} \cos^2(2\theta_0)\sin(2\phi_0) \right) \right]$$
$$+ \varepsilon_{xy} \left[ b_2 \left( 2b_2\widetilde{\chi}_{12} \sin\theta_0 \sin\phi_0 \left( \sin^2\theta_0 \cos(2\phi_0) - \cos^2\theta_0(2\cos(2\phi_0) + 1) \right) - 2b_2\widetilde{\chi}_{21} \sin\theta_0 \sin\phi_0 \left( 2\cos^2\theta_0 \cos^2\phi_0 + \cos(2\theta_0)\cos(2\phi_0) \right) \right. \right.$$
$$\left. \left. -2b_2\widetilde{\chi}_{11} \sin(4\theta_0)\sin^2\phi_0 \cos\phi_0 - 2b_2\widetilde{\chi}_{22} \sin(2\theta_0)\cos\phi_0 \cos(2\phi_0) \right) \right] ,$$

$$\text{(S18)}$$

$$\sigma_{xz}^{\mathrm{me}} =$$

$$\left[ b_2 \sin\theta_0 \cos\theta_0 \cos\phi_0 + b_2 \widetilde{H}_1 \left( \widetilde{\chi}_{11} \cos(2\theta_0) \cos\phi_0 - \widetilde{\chi}_{21} \cos\theta_0 \sin\phi_0 \right) + b_2 \widetilde{H}_2 \left( \widetilde{\chi}_{12} \cos(2\theta_0) \cos\phi_0 - \widetilde{\chi}_{22} \cos\theta_0 \sin\phi_0 \right) \right]$$

$$+ \varepsilon_{xx} \left[ b_1 b_2 \cos\phi_0 \left( \sin\theta_0 \sin\phi_0 \left( \widetilde{\chi}_{12}(3\cos(2\theta_0) + 1) \cos\phi_0 + \widetilde{\chi}_{21}(3\cos(2\theta_0) + 1) \cos\phi_0 - 4\widetilde{\chi}_{22} \cos\theta_0 \sin\phi_0 \right) - \widetilde{\chi}_{11} \sin(4\theta_0) \cos^2\phi_0 \right) \right]$$

$$+ \varepsilon_{yy} \left[ b_1 b_2 \left( \widetilde{\chi}_{22} \sin(2\theta_0) \sin(2\phi_0) \sin\phi_0 + 2\widetilde{\chi}_{21} \sin\theta_0 \sin\phi_0 \left( \cos^2\theta_0 \sin^2\phi_0 - \cos(2\theta_0) \cos^2\phi_0 \right) + \widetilde{\chi}_{12} \sin\theta_0 \left( \cos^2\theta_0 (\sin\phi_0 - \sin(3\phi_0)) \right. \right. \right.$$
$$\left. \left. \left. + \sin^2\theta_0 \sin(2\phi_0) \cos\phi_0 \right) + \widetilde{\chi}_{11} \sin(4\theta_0) \sin^2\phi_0(-\cos\phi_0) \right) \right]$$

$$+ \varepsilon_{zz} \left[ b_1 b_2 \left( \widetilde{\chi}_{11} \sin(4\theta_0) \cos\phi_0 - 2 \left( \widetilde{\chi}_{12} + \widetilde{\chi}_{21} \right) \sin\theta_0 \cos^2\theta_0 \sin\phi_0 \right) \right]$$

$$+ \varepsilon_{yz} \left[ b_2 \left( -2b_2 \widetilde{\chi}_{12} \cos(2\theta_0) \cos\theta_0 \cos(2\phi_0) + b_2 \widetilde{\chi}_{21}(-\cos\theta_0 - \cos(3\theta_0)) \cos(2\phi_0) + 4b_2 \widetilde{\chi}_{22} \cos^2\theta_0 \sin\phi_0 \cos\phi_0 \right. \right.$$
$$\left. \left. - 2b_2 \widetilde{\chi}_{11} \cos^2(2\theta_0) \sin(2\phi_0) \right) \right]$$

$$+ \varepsilon_{xz} \left[ b_2 \left( -4b_2 \widetilde{\chi}_{11} \cos^2(2\theta_0) \cos^2\phi_0 - 4b_2 \widetilde{\chi}_{22} \cos^2\theta_0 \sin^2\phi_0 + b_2 \widetilde{\chi}_{12}(\cos\theta_0 + \cos(3\theta_0)) \sin(2\phi_0) + b_2 \widetilde{\chi}_{21}(\cos\theta_0 + \cos(3\theta_0)) \sin(2\phi_0) \right) \right]$$

$$+ \varepsilon_{xy} \left[ b_2 \left( 2b_2 \widetilde{\chi}_{12} \sin\theta_0 \cos\phi_0 \left( \cos^2\theta_0 (1 - 2\cos(2\phi_0)) + \sin^2\theta_0 \cos(2\phi_0) \right) - 2b_2 \widetilde{\chi}_{21} \sin\theta_0 \cos\phi_0 \left( \cos(2\theta_0) \cos(2\phi_0) - 2\cos^2\theta_0 \sin^2\phi_0 \right) \right. \right.$$
$$\left. \left. - 2b_2 \widetilde{\chi}_{11} \sin(4\theta_0) \sin\phi_0 \cos^2\phi_0 + 2b_2 \widetilde{\chi}_{22} \sin(2\theta_0) \sin\phi_0 \cos(2\phi_0) \right) \right], \tag{S19}$$

$$\sigma_{xy}^{\mathrm{me}} =$$

$$\left[ b_2 \sin^2\theta_0 \sin\phi_0 \cos\phi_0 + b_2 \widetilde{H}_1 \sin\theta_0 \left( \widetilde{\chi}_{11} \cos\theta_0 \sin(2\phi_0) + \widetilde{\chi}_{21} \cos(2\phi_0) \right) + b_2 \widetilde{H}_2 \sin\theta_0 \left( \widetilde{\chi}_{12} \cos\theta_0 \sin(2\phi_0) + \widetilde{\chi}_{22} \cos(2\phi_0) \right) \right]$$

$$+ \varepsilon_{xx} \left[ -2b_1 b_2 \sin^2\theta_0 \cos\phi_0 \left( \widetilde{\chi}_{12} \cos\theta_0 \cos(3\phi_0) + \widetilde{\chi}_{21} \cos\theta_0 \cos(3\phi_0) + 4\widetilde{\chi}_{11} \cos^2\theta_0 \sin\phi_0 \cos^2\phi_0 + \widetilde{\chi}_{22}(\sin\phi_0 - \sin(3\phi_0)) \right) \right]$$

$$+ \varepsilon_{yy} \left[ -b_1 b_2 \sin\theta_0 \left( \widetilde{\chi}_{22} \sin\theta_0 \sin(4\phi_0) + 8\widetilde{\chi}_{11} \sin\theta_0 \cos^2\theta_0 \sin^3\phi_0 \cos\phi_0 + \widetilde{\chi}_{12} \sin(2\theta_0) \sin^2\phi_0 (2\cos(2\phi_0) + 1) \right. \right.$$
$$\left. \left. + \widetilde{\chi}_{21} \sin(2\theta_0) \sin^2\phi_0 (2\cos(2\phi_0) + 1)) \right) \right]$$

$$+ \varepsilon_{zz} \left[ b_1 b_2 \sin\theta_0 \left( 8\widetilde{\chi}_{11} \sin\theta_0 \cos^2\theta_0 \sin\phi_0 \cos\phi_0 + (\widetilde{\chi}_{12} + \widetilde{\chi}_{21}) \sin(2\theta_0) \cos(2\phi_0) \right) \right]$$

$$+ \varepsilon_{yz} \left[ b_2 \sin\theta_0 \left( -4b_2 \widetilde{\chi}_{22} \cos\theta_0 \cos\phi_0 \cos(2\phi_0) - 2b_2 \widetilde{\chi}_{12} \sin\phi_0 \left( \cos^2\theta_0 (2\cos(2\phi_0) + 1) - \sin^2\theta_0 \cos(2\phi_0) \right) \right. \right.$$
$$\left. \left. - 2b_2 \widetilde{\chi}_{21} \sin\phi_0 \left( 2\cos^2\theta_0 \cos(2\phi_0) + \cos(2\theta_0) \cos(2\phi_0) \right) - 8b_2 \widetilde{\chi}_{11} \cos(2\theta_0) \sin^2\phi_0 \cos\theta_0 \cos\phi_0 \right) \right]$$

$$+ \varepsilon_{xz} \left[ b_2 \sin\theta_0 \left( 2b_2 \widetilde{\chi}_{12} \cos\phi_0 \left( \cos^2\theta_0 (1 - 2\cos(2\phi_0)) + \sin^2\theta_0 \cos(2\phi_0) \right) - 2b_2 \widetilde{\chi}_{21} \sin\phi_0 \cos\phi_0 \left( \cos(2\theta_0) \cos(2\phi_0) - 2\cos^2\theta_0 \sin^2\phi_0 \right) \right. \right.$$
$$\left. \left. - 8b_2 \widetilde{\chi}_{11} \cos\theta_0 \cos(2\theta_0) \sin\phi_0 \cos^2\phi_0 + 4b_2 \widetilde{\chi}_{22} \cos\theta_0 \sin\phi_0 \cos(2\phi_0) \right) \right]$$

$$+ \varepsilon_{xy} \left[ b_2 \sin\theta_0 \left( -b_2 \widetilde{\chi}_{12} \sin(2\theta_0) \sin(4\phi_0) - b_2 \widetilde{\chi}_{21} \sin(2\theta_0) \sin(4\phi_0) - 4b_2 \widetilde{\chi}_{11} \sin\theta_0 \cos^2\theta_0 \sin^2(2\phi_0) - 4b_2 \widetilde{\chi}_{22} \sin\theta_0 \cos^2(2\phi_0) \right) \right]. \tag{S20}$$

The reduced Hooke's law Eq.(S13) says

$$\sigma_i^{\mathrm{me}} = C_{ij}^{\mathrm{me}} \varepsilon_j, \tag{S21}$$

and we can obtain the all 36 components of the second-rank stiffness tensor $C_{ij}^{\mathrm{me}}$ induced by magneto-elastic coupling by combining Eq.(S15) to Eq.(S20):

$$C_{11}^{\mathrm{me}} = -2b_1^2 \sin\theta_0 \left(\widetilde{\chi}_{22}\sin\theta_0\sin^2(2\phi_0) - 2\widetilde{\chi}_{12}\sin(2\theta_0)\sin\phi_0\cos^3\phi_0\right.$$
$$\left. -2\widetilde{\chi}_{21}\sin(2\theta_0)\sin\phi_0\cos^3\phi_0 + 4\widetilde{\chi}_{11}\sin\theta_0\cos^2\theta_0\cos^4\phi_0\right),$$
(S22-1)

$$C_{12}^{\mathrm{me}} = -\frac{1}{2}b_1^2\sin\theta_0\left(-4\widetilde{\chi}_{22}\sin\theta_0\sin^2(2\phi_0) + \widetilde{\chi}_{12}\sin(2\theta_0)\sin(4\phi_0) + \widetilde{\chi}_{21}\sin(2\theta_0)\sin(4\phi_0) + 4\widetilde{\chi}_{11}\sin\theta_0\cos^2\theta_0\sin^2(2\phi_0)\right),$$
(S22-2)

$$C_{13}^{\mathrm{me}} = -4b_1^2\sin^2\theta_0\cos\theta_0\cos\phi_0\left((\widetilde{\chi}_{12}+\widetilde{\chi}_{21})\sin\phi_0 - 2\widetilde{\chi}_{11}\cos\theta_0\cos\phi_0\right),$$
(S22-3)

$$C_{14}^{\mathrm{me}} = \frac{1}{2}b_1\sin\theta_0\left(-2b_2\widetilde{\chi}_{12}\left(\sin^2\theta_0\sin\phi_0\sin(2\phi_0)+\cos^2\theta_0(\cos\phi_0+\cos(3\phi_0))\right) - 8b_2\widetilde{\chi}_{11}\cos\theta_0\cos(2\theta_0)\sin\phi_0\cos^2\phi_0\right.$$
$$\left. +8b_2\widetilde{\chi}_{22}\cos\theta_0\sin\phi_0\cos^2\phi_0 + b_2\widetilde{\chi}_{21}\left(4\cos(2\theta_0)\sin^2\phi_0\cos\phi_0 - 4\cos^2\theta_0\cos^3\phi_0\right)\right),$$
(S22-4)

$$C_{15}^{\mathrm{me}} = \frac{1}{2}b_1\sin\theta_0\left(-8b_2\widetilde{\chi}_{11}\cos\theta_0\cos(2\theta_0)\cos^3\phi_0 + 2b_2\widetilde{\chi}_{12}(3\cos(2\theta_0)+1)\sin\phi_0\cos^2\phi_0 + 2b_2\widetilde{\chi}_{21}(3\cos(2\theta_0)+1)\sin\phi_0\cos^2\phi_0\right.$$
$$\left. -8b_2\widetilde{\chi}_{22}\cos\theta_0\sin^2\phi_0\cos\phi_0\right)$$
(S22-5)

$$C_{16}^{\mathrm{me}} = \frac{1}{2}b_1\sin\theta_0\left(2b_2\widetilde{\chi}_{22}\sin\theta_0\sin(4\phi_0) + 2b_2\widetilde{\chi}_{12}\sin(2\theta_0)(1-2\cos(2\phi_0))\cos^2\phi_0 + 2b_2\widetilde{\chi}_{21}\sin(2\theta_0)(1-2\cos(2\phi_0))\cos^2\phi_0\right.$$
$$\left. -16b_2\widetilde{\chi}_{11}\sin\theta_0\cos^2\theta_0\sin\phi_0\cos^3\phi_0\right),$$
(S22-6)

$$C_{21}^{\mathrm{me}} = -\frac{1}{2}b_1^2\sin\theta_0\left(-4\widetilde{\chi}_{22}\sin\theta_0\sin^2(2\phi_0) + \widetilde{\chi}_{12}\sin(2\theta_0)\sin(4\phi_0) + \widetilde{\chi}_{21}\sin(2\theta_0)\sin(4\phi_0) + 4\widetilde{\chi}_{11}\sin\theta_0\cos^2\theta_0\sin^2(2\phi_0)\right),$$
(S22-7)

$$C_{22}^{\mathrm{me}} = -2b_1^2\sin\theta_0\left(\widetilde{\chi}_{22}\sin\theta_0\sin^2(2\phi_0) + 4\widetilde{\chi}_{11}\sin\theta_0\cos^2\theta_0\sin^4\phi_0 + 2\widetilde{\chi}_{12}\sin(2\theta_0)\sin^3\phi_0\cos\phi_0 + 2\widetilde{\chi}_{21}\sin(2\theta_0)\sin^3\phi_0\cos\phi_0\right),$$
(S22-8)

$$C_{23}^{\mathrm{me}} = 2b_1^2\sin\theta_0\sin(2\theta_0)\sin\phi_0\left(2\widetilde{\chi}_{11}\cos\theta_0\sin\phi_0 + (\widetilde{\chi}_{12}+\widetilde{\chi}_{21})\cos\phi_0\right),$$
(S22-9)

$$C_{24}^{\mathrm{me}} = \frac{1}{2}b_1\sin\theta_0\left(-8b_2\widetilde{\chi}_{22}\cos\theta_0\sin\phi_0\cos^2\phi_0 - 8b_2\widetilde{\chi}_{11}\cos\theta_0\cos(2\theta_0)\sin^3\phi_0 - 2b_2\widetilde{\chi}_{12}(3\cos(2\theta_0)+1)\sin^2\phi_0\cos\phi_0\right.$$
$$\left. -2b_2\widetilde{\chi}_{21}(3\cos(2\theta_0)+1)\sin^2\phi_0\cos\phi_0\right),$$
(S22-10)

$$C_{25}^{\mathrm{me}} = \frac{1}{2}b_1\sin\theta_0\left(b_2\widetilde{\chi}_{21}\left(4\cos^2\theta_0\sin^3\phi_0 - 4\cos(2\theta_0)\sin\phi_0\cos^2\phi_0\right) + b_2\widetilde{\chi}_{12}\left(4\sin^2\theta_0\sin\phi_0\cos^2\phi_0 + 2\cos^2\theta_0(\sin\phi_0-\sin(3\phi_0))\right)\right.$$
$$\left. -8b_2\widetilde{\chi}_{11}\cos\theta_0\cos(2\theta_0)\sin^2\phi_0\cos\phi_0 + 8b_2\widetilde{\chi}_{22}\cos\theta_0\sin^2\phi_0\cos\phi_0\right),$$
(S22-11)

$$C_{26}^{\mathrm{me}} = \frac{1}{2}b_1\sin\theta_0\left(-2b_2\widetilde{\chi}_{22}\sin\theta_0\sin(4\phi_0) - 16b_2\widetilde{\chi}_{11}\sin\theta_0\cos^2\theta_0\sin^3\phi_0\cos\phi_0 - 2b_2\widetilde{\chi}_{12}\sin(2\theta_0)\sin^2\phi_0(2\cos(2\phi_0)+1)\right.$$
$$\left. -2b_2\widetilde{\chi}_{21}\sin(2\theta_0)\sin^2\phi_0(2\cos(2\phi_0)+1)\right),$$
(S22-12)

$$C_{31}^{\mathrm{me}} = 2b_1^2\sin^2\theta_0\cos\theta_0\left(4\widetilde{\chi}_{11}\cos\theta_0\cos^2\phi_0 - (\widetilde{\chi}_{12}+\widetilde{\chi}_{21})\sin(2\phi_0)\right),$$
(S22-13)

$$C_{32}^{\mathrm{me}} = 2b_1^2\sin^2\theta_0\cos\theta_0\left(4\widetilde{\chi}_{11}\cos\theta_0\sin^2\phi_0 + (\widetilde{\chi}_{12}+\widetilde{\chi}_{21})\sin(2\phi_0)\right),$$
(S22-14)

$$C_{33}^{\mathrm{me}} = -8b_1^2\widetilde{\chi}_{11}\sin^2\theta_0\cos^2\theta_0,$$
(S22-15)

$$C_{34}^{\mathrm{me}} = \frac{1}{2}b_1\cos\theta_0\left(8b_2\widetilde{\chi}_{11}\sin\theta_0\cos(2\theta_0)\sin\phi_0 + 4b_2\widetilde{\chi}_{12}\sin\theta_0\cos\theta_0\cos\phi_0 + 4b_2\widetilde{\chi}_{21}\sin\theta_0\cos\theta_0\cos\phi_0\right),$$
(S22-16)

$$C_{35}^{\mathrm{me}} = \frac{1}{2}b_1\cos\theta_0\left(8b_2\widetilde{\chi}_{11}\sin\theta_0\cos(2\theta_0)\cos\phi_0 - 4b_2\widetilde{\chi}_{12}\sin\theta_0\cos\theta_0\sin\phi_0 - 4b_2\widetilde{\chi}_{21}\sin\theta_0\cos\theta_0\sin\phi_0\right),$$
(S22-17)

$$C_{36}^{\mathrm{me}} = \frac{1}{2}b_1\cos\theta_0\left(8b_2\widetilde{\chi}_{11}\sin^2\theta_0\cos\theta_0\sin(2\phi_0) + 4b_2\widetilde{\chi}_{12}\sin^2\theta_0\cos(2\phi_0) + 4b_2\widetilde{\chi}_{21}\sin^2\theta_0\cos(2\phi_0)\right),$$
(S22-18)

$$C_{41}^{\mathrm{me}} = b_1b_2\left(\widetilde{\chi}_{11}\sin(4\theta_0)\sin\phi_0\left(-\cos^2\phi_0\right) - \sin\theta_0\left(\widetilde{\chi}_{12}\left(\sin^2\theta_0\sin\phi_0\sin(2\phi_0)+\cos^2\theta_0(\cos\phi_0+\cos(3\phi_0))\right)\right.\right.$$
$$\left.\left. -4\widetilde{\chi}_{22}\cos\theta_0\sin\phi_0\cos^2\phi_0 + \widetilde{\chi}_{21}\left(2\cos^2\theta_0\cos^3\phi_0 - 2\cos(2\theta_0)\sin^2\phi_0\cos\phi_0\right)\right)\right),$$
(S22-19)

$$C_{42}^{\mathrm{me}} = -b_1b_2\sin\phi_0\left(\widetilde{\chi}_{11}\sin(4\theta_0)\sin^2\phi_0 + \sin\theta_0\cos\phi_0\left(4\widetilde{\chi}_{22}\cos\theta_0\cos\phi_0 + \widetilde{\chi}_{12}(3\cos(2\theta_0)+1)\sin\phi_0 + \widetilde{\chi}_{21}(3\cos(2\theta_0)+1)\sin\phi_0\right)\right),$$
(S22-20)

$$C_{43}^{\mathrm{me}} = b_1b_2\left(\widetilde{\chi}_{11}\sin(4\theta_0)\sin\phi_0 + 2\left(\widetilde{\chi}_{12}+\widetilde{\chi}_{21}\right)\sin\theta_0\cos^2\theta_0\cos\phi_0\right),$$
(S22-21)

$$C_{44}^{\mathrm{me}} = \frac{1}{2}b_2\left(-4b_2\widetilde{\chi}_{22}\cos^2\theta_0\cos^2\phi_0 - 4b_2\widetilde{\chi}_{11}\cos^2(2\theta_0)\sin^2\phi_0 - 4b_2\widetilde{\chi}_{21}\cos\theta_0\cos(2\theta_0)\sin\phi_0\cos\phi_0 - 2b_2\widetilde{\chi}_{12}\cos\theta_0\cos(2\theta_0)\sin(2\phi_0)\right),$$
(S22-22)

$$C_{45}^{\mathrm{me}} = \frac{1}{2}b_2\left(-2b_2\widetilde{\chi}_{12}\cos(2\theta_0)\cos\theta_0\cos(2\phi_0) + b_2\widetilde{\chi}_{21}(-\cos\theta_0-\cos(3\theta_0))\cos(2\phi_0) + 4b_2\widetilde{\chi}_{22}\cos^2\theta_0\sin\phi_0\cos\phi_0\right.$$
$$\left. -2b_2\widetilde{\chi}_{11}\cos^2(2\theta_0)\sin(2\phi_0)\right),$$
(S22-23)

$$C_{46}^{\mathrm{me}} = \frac{1}{2}b_2\left(2b_2\widetilde{\chi}_{12}\sin\theta_0\sin\phi_0\left(\sin^2\theta_0\cos(2\phi_0)-\cos^2\theta_0(2\cos(2\phi_0)+1)\right) - 2b_2\widetilde{\chi}_{21}\sin\theta_0\sin\phi_0\left(2\cos^2\theta_0\cos^2\phi_0 + \cos(2\theta_0)\cos(2\phi_0)\right)\right.$$

$$-2b_2\widetilde{\chi}_{11}\sin(4\theta_0)\sin^2\phi_0\cos\phi_0-2b_2\widetilde{\chi}_{22}\sin(2\theta_0)\cos\phi_0\cos(2\phi_0))\,,\tag{S22-24}$$

$$C_{51}^{\mathrm{me}}=b_1b_2\cos\phi_0\left(\sin\theta_0\sin\phi_0\left(\widetilde{\chi}_{12}(3\cos(2\theta_0)+1)\cos\phi_0+\widetilde{\chi}_{21}(3\cos(2\theta_0)+1)\cos\phi_0-4\widetilde{\chi}_{22}\cos\theta_0\sin\phi_0\right)-\widetilde{\chi}_{11}\sin(4\theta_0)\cos^2\phi_0\right)\,,\tag{S22-25}$$

$$C_{52}^{\mathrm{me}}=b_1b_2\left(\widetilde{\chi}_{22}\sin(2\theta_0)\sin(2\phi_0)\sin\phi_0+2\widetilde{\chi}_{21}\sin\theta_0\sin\phi_0\left(\cos^2\theta_0\sin^2\phi_0-\cos(2\theta_0)\cos^2\phi_0\right)+\widetilde{\chi}_{12}\sin\theta_0\left(\cos^2\theta_0(\sin\phi_0-\sin(3\phi_0))\right.\right.$$
$$\left.\left.+\sin^2\theta_0\sin(2\phi_0)\cos\phi_0\right)+\widetilde{\chi}_{11}\sin(4\theta_0)\sin^2\phi_0(-\cos\phi_0)\right)\,,\tag{S22-26}$$

$$C_{53}^{\mathrm{me}}=b_1b_2\left(\widetilde{\chi}_{11}\sin(4\theta_0)\cos\phi_0-2\left(\widetilde{\chi}_{12}+\widetilde{\chi}_{21}\right)\sin\theta_0\cos^2\theta_0\sin\phi_0\right)\,,\tag{S22-27}$$

$$C_{54}^{\mathrm{me}}=\frac{1}{2}b_2\left(-2b_2\widetilde{\chi}_{12}\cos(2\theta_0)\cos\theta_0\cos(2\phi_0)+b_2\widetilde{\chi}_{21}(-\cos\theta_0-\cos(3\theta_0))\cos(2\phi_0)+4b_2\widetilde{\chi}_{22}\cos^2\theta_0\sin\phi_0\cos\phi_0\right.$$
$$\left.-2b_2\widetilde{\chi}_{11}\cos^2(2\theta_0)\sin(2\phi_0)\right)\,,\tag{S22-28}$$

$$C_{55}^{\mathrm{me}}=\frac{1}{2}b_2\left(-4b_2\widetilde{\chi}_{11}\cos^2(2\theta_0)\cos^2\phi_0-4b_2\widetilde{\chi}_{22}\cos^2\theta_0\sin^2\phi_0+b_2\widetilde{\chi}_{12}(\cos\theta_0+\cos(3\theta_0))\sin(2\phi_0)+b_2\widetilde{\chi}_{21}(\cos\theta_0+\cos(3\theta_0))\sin(2\phi_0)\right)\,,\tag{S22-29}$$

$$C_{56}^{\mathrm{me}}=\frac{1}{2}b_2\left(2b_2\widetilde{\chi}_{12}\sin\theta_0\cos\phi_0\left(\cos^2\theta_0(1-2\cos(2\phi_0))+\sin^2\theta_0\cos(2\phi_0)\right)-2b_2\widetilde{\chi}_{21}\sin\theta_0\cos\phi_0\left(\cos(2\theta_0)\cos(2\phi_0)-2\cos^2\theta_0\sin^2\phi_0\right)\right.$$
$$\left.-2b_2\widetilde{\chi}_{11}\sin(4\theta_0)\sin\phi_0\cos^2\phi_0+2b_2\widetilde{\chi}_{22}\sin(2\theta_0)\sin\phi_0\cos(2\phi_0)\right)\,,\tag{S22-30}$$

$$C_{61}^{\mathrm{me}}=-2b_1b_2\sin^2\theta_0\cos\phi_0\left(\widetilde{\chi}_{12}\cos\theta_0\cos(3\phi_0)+\widetilde{\chi}_{21}\cos\theta_0\cos(3\phi_0)+4\widetilde{\chi}_{11}\cos^2\theta_0\sin\phi_0\cos^2\phi_0+\widetilde{\chi}_{22}(\sin\phi_0-\sin(3\phi_0))\right)\,,\tag{S22-31}$$

$$C_{62}^{\mathrm{me}}=-b_1b_2\sin\theta_0\left(\widetilde{\chi}_{22}\sin\theta_0\sin(4\phi_0)+8\widetilde{\chi}_{11}\sin\theta_0\cos^2\theta_0\sin^3\phi_0\cos\phi_0+\widetilde{\chi}_{12}\sin(2\theta_0)\sin^2\phi_0(2\cos(2\phi_0)+1)\right.$$
$$\left.+\widetilde{\chi}_{21}\sin(2\theta_0)\sin^2\phi_0(2\cos(2\phi_0)+1)\right)\,,\tag{S22-32}$$

$$C_{63}^{\mathrm{me}}=b_1b_2\sin\theta_0\left(8\widetilde{\chi}_{11}\sin\theta_0\cos^2\theta_0\sin\phi_0\cos\phi_0+(\widetilde{\chi}_{12}+\widetilde{\chi}_{21})\sin(2\theta_0)\cos(2\phi_0)\right)\,,\tag{S22-33}$$

$$C_{64}^{\mathrm{me}}=\frac{1}{2}b_2\left(-2b_2\widetilde{\chi}_{12}\cos(2\theta_0)\cos\theta_0\cos(2\phi_0)+b_2\widetilde{\chi}_{21}(-\cos\theta_0-\cos(3\theta_0))\cos(2\phi_0)+4b_2\widetilde{\chi}_{22}\cos^2\theta_0\sin\phi_0\cos\phi_0\right.$$
$$\left.-2b_2\widetilde{\chi}_{11}\cos^2(2\theta_0)\sin(2\phi_0)\right)\,,\tag{S22-34}$$

$$C_{65}^{\mathrm{me}}=\frac{1}{2}b_2\sin\theta_0\left(2b_2\widetilde{\chi}_{12}\cos\phi_0\left(\cos^2\theta_0(1-2\cos(2\phi_0))+\sin^2\theta_0\cos(2\phi_0)\right)-2b_2\widetilde{\chi}_{21}\cos\phi_0\left(\cos(2\theta_0)\cos(2\phi_0)-2\cos^2\theta_0\sin^2\phi_0\right)\right.$$
$$\left.-8b_2\widetilde{\chi}_{11}\cos\theta_0\cos(2\theta_0)\sin\phi_0\cos^2\phi_0+4b_2\widetilde{\chi}_{22}\cos\theta_0\sin\phi_0\cos(2\phi_0)\right)\,,\tag{S22-35}$$

$$C_{66}^{\mathrm{me}}=\frac{1}{2}b_2\sin\theta_0\left(-b_2\widetilde{\chi}_{12}\sin(2\theta_0)\sin(4\phi_0)-b_2\widetilde{\chi}_{21}\sin(2\theta_0)\sin(4\phi_0)-4b_2\widetilde{\chi}_{11}\sin\theta_0\cos^2\theta_0\sin^2(2\phi_0)-4b_2\widetilde{\chi}_{22}\sin\theta_0\cos^2(2\phi_0)\right)\,.\tag{S22-36}$$

Due to symmetry, the stiffness tensor $C_{ij}^{\mathrm{me}}$ should be symmetric [5, 6] since $C_{ij}^{\mathrm{me}}=\frac{\partial\sigma_i^{\mathrm{me}}}{\partial\varepsilon_j}=\frac{\partial^2\mathcal{U}_{\mathrm{me}}}{\partial\varepsilon_i\partial\varepsilon_j}$, and $\frac{\partial^2\mathcal{U}_{\mathrm{me}}}{\partial\varepsilon_i\partial\varepsilon_j}=\frac{\partial^2\mathcal{U}_{\mathrm{me}}}{\partial\varepsilon_j\partial\varepsilon_i}$. However, the effective tensor $C_{ij}^{\mathrm{me}}$ derived above Eq.(S22) is *asymmetric*, *e.g.*, for certain components $i\neq j$, $C_{ij}^{\mathrm{me}}\neq C_{ji}^{\mathrm{me}}$. The reason is that we have kept only first-order terms during the derivation.

The effective stiffness tensor Eq.(S22) can be seemed as a correction induced by magneto-elastic coupling. In order to have a complete description on the elastic property of the magnetic material, we need to combine the original elastic stiffness tensor arising from elastic energy. The total effective stiffness tensor is expressed as

$$C_{ij}^{\mathrm{tot}}=C_{ij}^{\mathrm{el}}+C_{ij}^{\mathrm{me}}=\begin{pmatrix}2\mu+\lambda&\lambda&\lambda&0&0&0\\\lambda&2\mu+\lambda&\lambda&0&0&0\\\lambda&\lambda&2\mu+\lambda&0&0&0\\0&0&0&\mu&0&0\\0&0&0&0&\mu&0\\0&0&0&0&0&\mu\end{pmatrix}+\begin{pmatrix}C_{11}^{\mathrm{me}}&C_{12}^{\mathrm{me}}&C_{13}^{\mathrm{me}}&C_{14}^{\mathrm{me}}&C_{15}^{\mathrm{me}}&C_{16}^{\mathrm{me}}\\C_{21}^{\mathrm{me}}&C_{22}^{\mathrm{me}}&C_{23}^{\mathrm{me}}&C_{24}^{\mathrm{me}}&C_{25}^{\mathrm{me}}&C_{26}^{\mathrm{me}}\\C_{31}^{\mathrm{me}}&C_{32}^{\mathrm{me}}&C_{33}^{\mathrm{me}}&C_{34}^{\mathrm{me}}&C_{35}^{\mathrm{me}}&C_{36}^{\mathrm{me}}\\C_{41}^{\mathrm{me}}&C_{42}^{\mathrm{me}}&C_{43}^{\mathrm{me}}&C_{44}^{\mathrm{me}}&C_{45}^{\mathrm{me}}&C_{46}^{\mathrm{me}}\\C_{51}^{\mathrm{me}}&C_{52}^{\mathrm{me}}&C_{53}^{\mathrm{me}}&C_{54}^{\mathrm{me}}&C_{55}^{\mathrm{me}}&C_{56}^{\mathrm{me}}\\C_{61}^{\mathrm{me}}&C_{62}^{\mathrm{me}}&C_{63}^{\mathrm{me}}&C_{64}^{\mathrm{me}}&C_{65}^{\mathrm{me}}&C_{66}^{\mathrm{me}}\end{pmatrix}\,,\tag{S23}$$

with $\lambda$ and $\mu$ the Láme constants for the magnetic material. For non-magnetic materials, only the elastic stiffness tensor $C_{ij}^{\mathrm{el}}$ is considered.

From Eq.(S15) to Eq.(S20), there are stress terms which have no dependence on strain $\varepsilon_{ij}$ and cannot be included into the effective stiffness tensor Eq.(S22). These extra stress terms are from different physical origin. One set is constant stress,

$$\sigma_{xx}^0=b_1\sin^2\theta_0\cos^2\phi_0,\tag{S24-1}$$

$$\sigma_{yy}^0=b_1\sin^2\theta_0\sin^2\phi_0,\tag{S24-2}$$

$$\sigma_{zz}^0=b_1\cos^2\theta_0,\tag{S24-3}$$

$$\sigma_{yz}^0=b_2\sin\theta_0\cos\theta_0\sin\phi_0,\tag{S24-4}$$

$$\sigma_{xz}^0 = b_2 \sin\theta_0 \cos\theta_0 \cos\phi_0, \tag{S24-5}$$

$$\sigma_{xy}^0 = b_2 \sin^2\theta_0 \sin\phi_0 \cos\phi_0, \tag{S24-6}$$

which can be interpreted as magnetostriction, *i.e.*, static lattice deformation in the presence of static magnetization, consistent with [7]. Static magnetostriction is disregarded in numerical simulation since it doesn't contribute to dynamics in frequency domain. In reality, static deformation induced by magnetostriction may affect local crystalline anisotropy of the magnetic orders, which can be included as effective fields and doesn't affect the conclusion of this work.

Another set of effective stress arises from direct excitation by external fields,

$$\sigma_{xx}^{\mathrm{H}} = 2b_1\widetilde{H}_2 \sin\theta_0 \left(\widetilde{\chi}_{12}\cos\theta_0\cos^2\phi_0 - \widetilde{\chi}_{22}\sin\phi_0\cos\phi_0\right) + 2b_1\widetilde{H}_1\sin\theta_0\cos\phi_0\left(\widetilde{\chi}_{11}\cos\theta_0\cos\phi_0 - \widetilde{\chi}_{21}\sin\phi_0\right), \tag{S25-1}$$

$$\sigma_{yy}^{\mathrm{H}} = b_1\widetilde{H}_1 \sin\theta_0 \left(2\widetilde{\chi}_{11}\cos\theta_0\sin^2\phi_0 + \widetilde{\chi}_{21}\sin(2\phi_0)\right) + b_1\widetilde{H}_2\sin\theta_0\left(2\widetilde{\chi}_{12}\cos\theta_0\sin^2\phi_0 + \widetilde{\chi}_{22}\sin(2\phi_0)\right), \tag{S25-2}$$

$$\sigma_{zz}^{\mathrm{H}} = -2b_1\widetilde{H}_1\widetilde{\chi}_{11}\sin\theta_0\cos\theta_0 - 2b_1\widetilde{H}_2\widetilde{\chi}_{12}\sin\theta_0\cos\theta_0, \tag{S25-3}$$

$$\sigma_{yz}^{\mathrm{H}} = b_2\widetilde{H}_1\left(\widetilde{\chi}_{21}\cos\theta_0\cos\phi_0 + \widetilde{\chi}_{11}\cos(2\theta_0)\sin\phi_0\right) + b_2\widetilde{H}_2\left(\widetilde{\chi}_{22}\cos\theta_0\cos\phi_0 + \widetilde{\chi}_{12}\cos(2\theta_0)\sin\phi_0\right), \tag{S25-4}$$

$$\sigma_{xz}^{\mathrm{H}} = b_2\widetilde{H}_1\left(\widetilde{\chi}_{11}\cos(2\theta_0)\cos\phi_0 - \widetilde{\chi}_{21}\cos\theta_0\sin\phi_0\right) + b_2\widetilde{H}_2\left(\widetilde{\chi}_{12}\cos(2\theta_0)\cos\phi_0 - \widetilde{\chi}_{22}\cos\theta_0\sin\phi_0\right), \tag{S25-5}$$

$$\sigma_{xy}^{\mathrm{H}} = b_2\widetilde{H}_1\sin\theta_0\left(\widetilde{\chi}_{11}\cos\theta_0\sin(2\phi_0) + \widetilde{\chi}_{21}\cos(2\phi_0)\right) + b_2\widetilde{H}_2\sin\theta_0\left(\widetilde{\chi}_{12}\cos\theta_0\sin(2\phi_0) + \widetilde{\chi}_{22}\cos(2\phi_0)\right), \tag{S25-6}$$

which depends on magnitude and chirality of external field (dynamical component) $\widetilde{H}_1$ and $\widetilde{H}_2$, which are complex in general. Combining and Eq.(S25) and Eq.(S15-S20), we obtain Eq.(11) in the main text, which is generalized Hooke's law in the presence of magneto-elastic coupling as well as external microwave driving and the total stress is

$$\sigma_i^{\mathrm{tot}} = C_{ij}^{\mathrm{el}} + C_{ij}^{\mathrm{me}}\varepsilon_j + \sigma_i^{\mathrm{H}}, \qquad \text{(for magnets)} \tag{S26-1}$$

$$\sigma_i^{\mathrm{tot}} = C_{ij}^{\mathrm{el}}\varepsilon_j. \qquad \text{(for non-magnets)} \tag{S26-2}$$

The stress terms in Eq.(S26) (in Voigt notation) can be transformed into matrix notation

$$\bar{\boldsymbol{\sigma}} = \begin{pmatrix} \sigma_1 & \sigma_6 & \sigma_5 \\ \sigma_6 & \sigma_2 & \sigma_4 \\ \sigma_5 & \sigma_4 & \sigma_3 \end{pmatrix} = \begin{pmatrix} \sigma_{xx} & \sigma_{xy} & \sigma_{xz} \\ \sigma_{xy} & \sigma_{yy} & \sigma_{yz} \\ \sigma_{xz} & \sigma_{yz} & \sigma_{zz} \end{pmatrix}, \tag{S27}$$

and plugged into the equation of motion for elastic waves (Eq.(12) in the main text), which is solved numerically. It should be noticed that the generalized Hooke's law derived above is essentially a correction of the elastic stiffness tensor in frequency domain, which characterizes leading-order dynamic response to external perturbation and is different from the complete stiffness tensor obtained from linear magneto-elastic approximations reported in earlier literature [5].

## II. THEORETICAL STRENGTH OF COHERENT COUPLING AND DYNAMIC EXCHANGE COUPLING

Theoretical strength of coherent coupling and dynamic exchange coupling have been plotted in Fig.4 in the main text and are compared with numerical simulation.

Strength of coherent coupling between magnons and phonons is calculated according to overlap integral between the standing elastic waves and the Kittle mode confined in the magnetic layer [8–10]. Hence the coherent coupling between two magnets mediated by standing elastic waves at resonance can be expressed as

$$\Omega = \sqrt{N}b_2\sqrt{\frac{2\gamma}{\omega\rho_{\mathrm{M}}M_sLd}}\left(1 - \cos\frac{\omega d}{c_{\mathrm{M}}}\right), \tag{S28}$$

with $N = 2$ for two identical magnets, where only homogeneous Kittel mode is considered. By considering contribution from high-order inhomogeneous spin-wave modes, the coupling strength will get closer with the one extracted from simulation as well as measured in experiment [10].

Strength of dynamic exchange coupling is defined as Eq.(4) in the main text, where the effective Gilbert damping coefficient $\alpha'$ is induced by phonon pumping and has been analytically derived in [11]. For a magnet with magnetization along $\hat{x}$ attached on an infinite phonon sink, there is an effective field

$$\mathbf{H}_{\mathrm{me}} = \frac{b_2}{\mu_0 M_s}\mathrm{Re}(\nu)\hat{\mathbf{x}}, \tag{S29}$$

and an effective Gilbert damping

$$\alpha_{\mathrm{me}} = -\frac{\gamma b_2}{\omega \mu_0 M_s}\mathrm{Im}(\nu),$$ (S30)

with the dimensionless quantity

$$\nu = \frac{b_2}{\omega d \rho_{\mathrm{M}} c_{\mathrm{M}}}\frac{2\left[\cos\left(k_{\mathrm{M}} d\right) - 1\right] - i\frac{\rho_{\mathrm{NM}} c_{\mathrm{NM}}}{\rho_{\mathrm{M}} c_{\mathrm{M}}}\sin\left(k_{\mathrm{M}} d\right)}{\sin\left(k_{\mathrm{M}} d\right) + i\frac{\rho_{\mathrm{NM}} c_{\mathrm{NM}}}{\rho_{\mathrm{M}} c_{\mathrm{M}}}\cos\left(k_{\mathrm{M}} d\right)}.$$ (S31)

The effective field Eq.(S29) induces a frequency shift $\gamma H_{\mathrm{me}}/(2\pi) \sim 1\,\mathrm{MHz}$ on the Kittel mode and such frequency shift is observed in the simulation. The effective Gilbert damping is exactly what we define in the main text $\alpha' = \alpha_{\mathrm{me}}$, whose theoretical values are smaller than the simulated ones as shown in Fig.4(a) in the main text. There are two possible reasons: (i) The theoretical value is calculated with the assumption of an infinite phonon sink. However, the structure considered in this work is finite and reflection is inevitable even though large acoustic dissipation is assumed. (ii) The phonon absorption process is not purely reverse of phonon pumping so that $\alpha''$ is in general smaller than the theoretical evaluation, leading to overestimation of $\alpha'$. (iii) Similar to the coherent coupling, contribution from inhomogeneous spin-wave excitations is not included in the theoretical model while preserved in the numerical simulation.

Theoretical strength of both type of coupling are plotted in a wider frequency range in Fig. S1, with consideration of original elastic damping $\beta = \beta_0 = 2\eta_0 \rho_{\mathrm{NM}}$. Three specific cases are demonstrated: (i) Thickness $L = 0.5\,\mathrm{mm}$, which is a typical thickness performed in experiments [8, 12], (ii) Thickness $L = 5\,\mathrm{mm}$, the same as the one in Fig.(4) in the main text but with a wider frequency range, and (iii) Thickness $L = 10\,\mathrm{mm}$, much larger than the phonon decay length $\Lambda \simeq 2\,\mathrm{mm}$.

For case (i), strong coupling regime (white) covers a typical range of microwave measurement in laboratory, so that coherent coupling dominates and that's the reason why level attraction induced by dynamic exchange coupling has not been reported yet. For case (ii), strong coupling regime shrinks with increasing thickness. It's possible to observe level attraction in weak coupling regime, but the dynamic exchange coupling drops to almost zero at high frequencies, so only data in low-frequency regime are shown in Fig.(4) of the main text. For case (iii), the thickness is much larger than the phonon decay length, hence the elastic waves attenuates severely before reaching to another magnet, leading to almost zero strength for the dynamic exchange coupling in all frequency range.

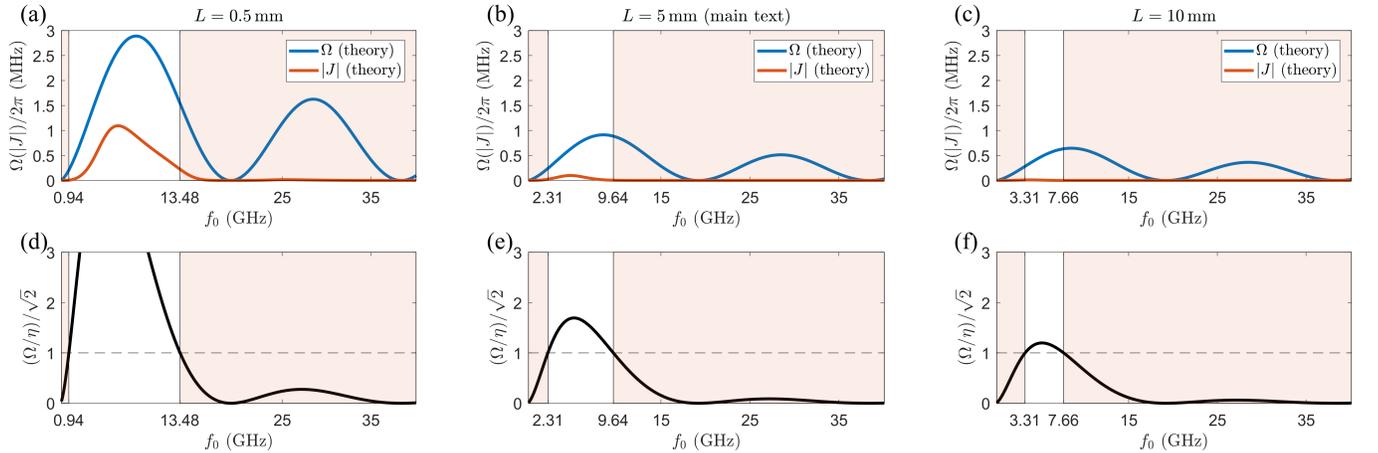

Figure S1. Frequency dependence of coherent coupling strength $\Omega$ (blue) and magnitude of dynamic exchange coupling strength $|J|$, and the relative strength between the coherent coupling rate and phonon relaxation rate $\Omega/(\sqrt{2}\eta)$. Original elastic damping $\beta = \beta_0$ of GGG is considered with thickness (a)(d) $L = 0.5\,\mathrm{mm}$, (b)(e) $L = 5\,\mathrm{mm}$ and (c)(f) $L = 10\,\mathrm{mm}$. Weak (strong) coupling regime is indicated in red (white).

## III. SPECTRUM TRANSITION FROM STRONG COUPLING REGIME TOWARDS WEAK COUPLING REGIME

In the main text (Fig.2), the case of perpendicular magnetization with $L = 0.5\,\text{mm}$ and amplified elastic damping $\beta = 6\beta_0$ is demonstrated. In this section, we show the other four cases with elastic damping ranging from $\beta = \beta_0$ (original damping measured in experiments) to $\beta = 10\beta_0$, as shown in Fig. S2.

When $\beta = \beta_0$ (Fig. S2(a,e)), the spectrum shows typical bright-mode splitting induced by coherent coupling, consistent with experimental observation [12], in which case $\Omega/(\sqrt{2}\eta) \simeq 3 > 1$ corresponding to strong coupling regime, so that the model for dynamic exchange coupling is invalid. The case $\beta = 4\beta_0$ (Fig. S2(b,f)) is intermediate, where $\Omega/(\sqrt{2}\eta) \simeq 1$ and features from both coherent coupling and dynamic exchange coupling are mixed. Although the theoretical curves do not agree with simulation well, the model successfully predict the position of exceptional points which evolves from shrinking central phonon modes. For the remaining two cases $\beta = 7\beta_0$ (Fig. S2(c,g)) and $\beta = 10\beta_0$ (Fig. S2(d,h)), the system moves to weak coupling regime and the proposed model is valid and agrees well with simulation.

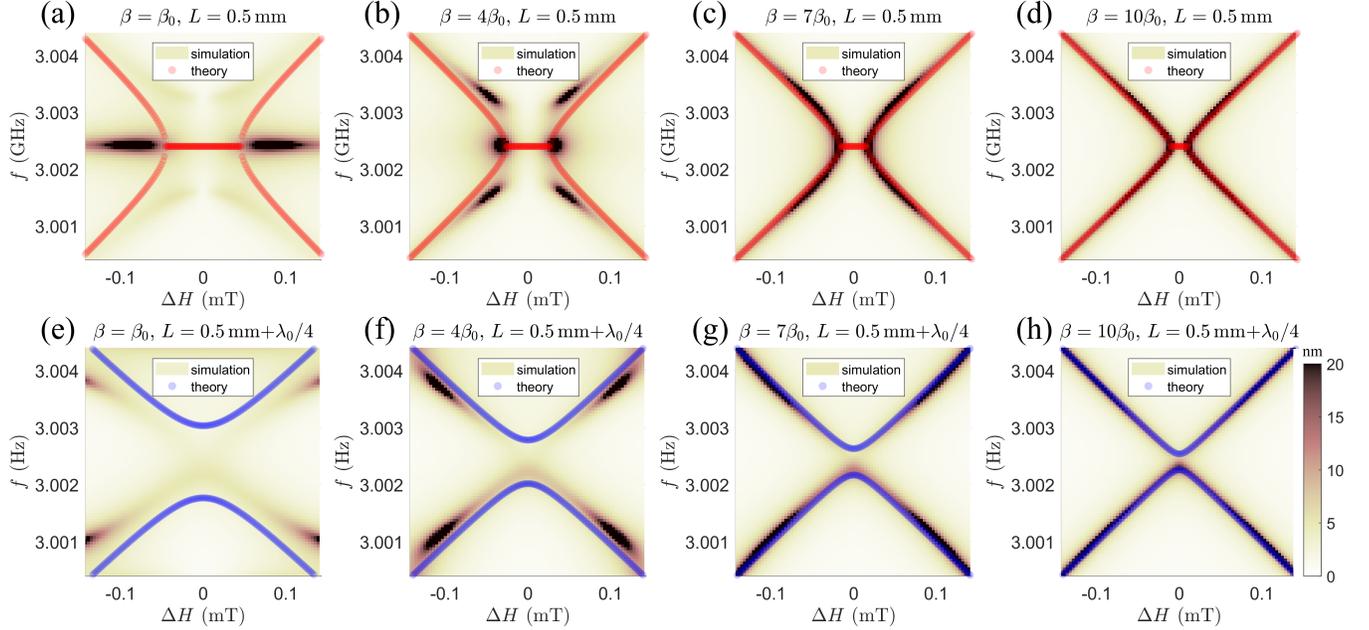

Figure S2. Phonon excitation spectrum for $L = 0.5\,\text{mm}$ (a-d) and $L = 0.5\,\text{mm}+\lambda_0/4$ (e-h) with different elastic damping. Theoretical curves for $\alpha' = 2.5 \times 10^{-4}$ are plotted for comparison.

[1] C. Kittel, Physical Review **110**, 836 (1958).

[2] L. Dreher, M. Weiler, M. Pernpeintner, H. Huebl, R. Gross, M. S. Brandt, and S. T. B. Goennenwein, Physical Review B **86**, 134415 (2012), publisher: American Physical Society.

[3] D. D. Stancil and A. Prabhakar, *Spin Waves: Theory and Applications* (Springer Science & Business Media, 2009).

[4] B. Lüthi, *Physical Acoustics in the Solid State* (Springer Science & Business Media, 2007).

[5] S. Rinaldi and G. Turilli, Physical Review B **31**, 3051 (1985).

[6] P. Nieves, J. Tranchida, S. Nikolov, A. Fraile, and D. Legut, Physical Review B **105**, 134430 (2022), publisher: American Physical Society.

[7] T. Sato, W. Yu, S. Streib, and G. E. W. Bauer, Physical Review B **104**, 014403 (2021), publisher: American Physical Society.

[8] K. An, A. N. Litvinenko, R. Kohno, A. A. Fuad, V. V. Naletov, L. Vila, U. Ebels, G. de Loubens, H. Hurdequint, N. Beaulieu, J. Ben Youssef, N. Vukadinovic, G. E. W. Bauer, A. N. Slavin, V. S. Tiberkevich, and O. Klein, Physical Review B **101**, 060407 (2020), publisher: American Physical Society.

[9] A. Litvinenko, R. Khymyn, V. Tyberkevych, V. Tikhonov, A. Slavin, and S. Nikitov, Physical Review Applied **15**, 034057 (2021), publisher: American Physical Society.

[10] R. Schlitz, L. Siegl, T. Sato, W. Yu, G. E. W. Bauer, H. Huebl, and S. T. B. Goennenwein, Physical Review B **106**, 014407 (2022), publisher: American Physical Society.

[11] S. Streib, H. Keshtgar, and G. E. Bauer, Physical Review Letters **121**, 027202 (2018).

[12] K. An, R. Kohno, A. Litvinenko, R. Seeger, V. Naletov, L. Vila, G. de Loubens, J. Ben Youssef, N. Vukadinovic, G. Bauer, A. Slavin, V. Tiberkevich, and O. Klein, Physical Review X **12**, 011060 (2022), publisher: American Physical Society.



\end{document}